\documentclass[11pt]{article}

\usepackage[preprint]{acl}

\usepackage{times}
\usepackage{latexsym}

\usepackage[T1]{fontenc}

\usepackage[utf8]{inputenc}

\usepackage{microtype}

\usepackage{inconsolata}

\usepackage{graphicx}

\usepackage{hyperref}       
\usepackage{url}            
\usepackage{booktabs}       
\usepackage{amsfonts}       
\usepackage{nicefrac}       
\usepackage{xcolor}         
\usepackage{amsmath}
\usepackage{bbding}
\usepackage{array}
\usepackage[caption=false,font=normalsize,labelfont=sf,textfont=sf]{subfig}
\usepackage{caption, subcaption}
\usepackage{textcomp}
\usepackage{stfloats}
\usepackage{verbatim}
\usepackage{multirow}
\usepackage[mathscr]{eucal}
\usepackage{bbm}
\usepackage{tabularx}
\usepackage[linesnumbered,ruled,vlined]{algorithm2e}
\usepackage[most]{tcolorbox}
\usepackage{ragged2e}
\usepackage{makecell}
\usepackage{pifont}
\usepackage{placeins}
\usepackage{natbib}
\usepackage{etoc}
\usepackage{lipsum}
\usepackage[table]{xcolor}
\usepackage{geometry}
\usepackage{amssymb}

\definecolor{lightyellow}{rgb}{1,1,0.8}
\definecolor{lightblue}{rgb}{0.8,0.9,1}
\hypersetup{
  colorlinks,               
  linkcolor={red!50!black}, 
  citecolor={blue!45!black},
  urlcolor=[HTML]{000099}
}

\title{MulFSA: Multi-level Financial Sentiment Analysis \\Framework for Bond Market}

\author{
 \textbf{Yiwei Liu\textsuperscript{1,3}},
 \textbf{Junbo Wang\textsuperscript{1}},
 \textbf{Lei Long\textsuperscript{1}},
 \textbf{Xin Li\textsuperscript{1}},
 \textbf{Ruiting Ma\textsuperscript{1}},
\\
 \textbf{Yuankai Wu\textsuperscript{1}},
 \textbf{Xuebin Chen\textsuperscript{1,2}},
\\
\\
 \textsuperscript{1}Sichuan University,
 \textsuperscript{2}Fudan University,
 \textsuperscript{3}National University of Singapore,
\\
 \small{
   \textbf{Correspondence:} \href{mailto:lew1sin@163.com}{lew1sin@163.com}, \href{mailto:chenxb@fudan.edu.cn}{chenxb@fudan.edu.cn}
 }
}

\begin{document}
\maketitle
\begin{abstract}
Existing financial sentiment analysis methods often fail to capture the multi-faceted nature of risk in bond markets due to their single-level approach and neglect of temporal dynamics. We propose Multi-level Financial Sentiment Analysis (MulFSA) based on pre-trained language models (PLMs) and large language models (LLMs), a novel framework that systematically integrates firm-specific micro-level sentiment, industry-specific meso-level sentiment, and duration-aware smoothing to model the latency and persistence of textual impact. Applying MulFSA to the comprehensive Chinese bond market corpus constructed by us (2013–2023, 1.35M texts), we extracted a daily composite sentiment index. Empirical results show statistically measurable improvements in credit spread forecasting when incorporating sentiment (10.25\% MAE and 11.94\% MAPE reduction), with sentiment shifts closely correlating with major social risk events and firm-specific crises. \textbf{Project Page}: \url{https://mulfsa.github.io/}. 

\end{abstract}

\section{Introduction}
\label{sec:intro} 

As an irrational shock, sentiment can influence financial markets by affecting asset prices, and other macroeconomic variables~\cite{perri2018international}. In financial quantitative research, financial sentiment analysis (FSA) aims to quantify investor sentiment (such as bullish, bearish, or neutral) from unstructured textual data and serves as a feature that complements traditional structured indices~\cite{aleti2024news}. By incorporating sentiment, forecasting models can be improved in downstream tasks such as forecasting trading volume, and risk~\cite{du2024financial,li2024extracting}.

FSA can be categorized into sentence-level sentiment analysis (SLSA) and aspect-based sentiment analysis (ABSA)~\cite{huang2020entity}. The former outputs the overall sentiment of a sentence, whereas the latter is treated as a sequence labeling problem that identifies the sentiment of a specific entity. The context of financial texts is often subtle and complex. Ahbali et al. \cite{ahbali2022identifying} pointed out that a single paragraph may contain different firms with different sentiments. For example, ``\textit{Tong Guang Convertible Bond surged over 11\%, ranking first in gains, while Tian Lu Convertible Bond led the decline, dropping more than 8\%.}'' Trivially, the sentiment polarities of \textit{Tong Guang} and \textit{Tian Lu} are opposite. But most existing FSA methods assume a single overall sentiment at the whole-document level, i.e., SLSA, without considering the heterogeneity of textual impact on different firms (firm-specific). Therefore, we argue that FSA requires a more fine-grained measurement, which leads to the first question:

\textbf{Q1}: How to precisely analyze the sentiment of entities with different polarities from a firm-specific view under subtle and complex contexts?

Financial texts may also describe industry environments, which can be regarded as industry-wide risks that may be transmitted to micro-level individuals~\cite{efretuei2021year}. Current FSA frameworks fail to consider the heterogeneity of impact on industries (industry-specific), and therefore cannot handle cases where a firm-specific sentiment deviates from the industry trend~\cite{huang2020entity}. This leads to the second question:

\textbf{Q2}: How to derive the sentiment transmitted from the broader environment to related firms from an industry-specific view?

Mechanisms such as \textit{Crash Narratives}~\cite{goetzmann2022crash} showed that the impact of financial texts on assets is non-instantaneous and nonlinear, i.e., the information diffuses into the market with latency and persistence~\cite{benhabib2016sentiments}. Current FSA frameworks fail to incorporate such duration, which leads to the third question:

\textbf{Q3}: How to capture the duration of text sentiment by considering its latency and persistence, enabling texts to interact with others and diffuse their effect to the entire time series?

Although pre-trained language models (PLMs) and large language models (LLMs) have gained increasing attention, many well-researched NLP tasks have not yet been grounded in financial applications. 
For Q1, it is a typical ABSA task, thus we finetuned a BERT to perform ABSA on multiple firms mentioned in the same text. For Q2, we first transform the text sentiment output by an LLM agent into a topic view, then use retrieval-augmented generation (RAG) based on a knowledge graph to propagate it to the relevant industries, and finally to the corresponding firms. For Q3, we aggregate firm-level and industry-level sentiment using an MLP and apply \textit{wavelet smoothing} to remove noise and extract a smooth long-term sentiment trend that captures interactions across texts. Therefore, our contributions are as follows:

\textbf{\textit{Data Collection}}: We chose the Chinese bond market as the subject, and constructed a text corpus spanning 2013–2023, which covers bond-related news, firm announcements, analyst reports, and firm disclosures, amounting to 1,353,585 entries. 

\textbf{\textit{Multi-Level Sentiment Extraction}}: We propose the first unified framework that integrates micro-level firm-specific sentiment, meso-level industry spillover effects, and the latency and persistence of textual impact, thereby mitigating the limitations of a single financial perspective.

\textbf{\textit{Bond Default Risk Forecasting Modeling}}: We selected bond default risk forecasting (BDRF) as the downstream task to backtest whether the extracted text sentiment can improve forecasting performance. Comparative and ablation experiments show a 10.25\% decrease in MAE and a 11.94\% decrease in MAPE, and that the extracted sentiment correlates with social risk events.

\section{Related Works}
\subsection{Financial Sentiment Analysis}
\label{sec:related_works_sa}
\begin{table}[htbp]
  \centering
  \tiny
  \setlength\tabcolsep{6.5pt}
    \caption{F1-score performance of FinLLMs (with 5-shot prompting) and FinPLMs on \textit{FPB} and \textit{FiQA-SA} benchmarks, with results from~\cite{xie2023pixiu,huang2024open,lee2025large}.}
  \begin{tabular}{lccc}
    \toprule
    \textbf{Model} & \textbf{FPB} & \textbf{FiQA-SA} & \textbf{Arch.} \\
    \midrule
    \textit{LLaMA3-8B}~\cite{llama3modelcard} & 0.69 & 0.52 & LLM \\
    \textit{BloombergGPT}~\cite{wu2023bloomberggpt} & 0.51 & 0.75 & LLM \\
    \textit{FinBERT (2019)}~\cite{araci2019finbert} & 0.90 & 0.63 & PLM \\
    \textit{FinBERT (2022)}~\cite{huang2023finbert} & 0.89 & 0.65 & PLM \\
    \textit{FinLLaMA}~\cite{huang2024open} & 0.70 & 0.75 & LLM \\
    \textit{GPT-4}~\cite{achiam2023gpt} & 0.78 & 0.80 & LLM \\
    \textit{FinMA-7B}~\cite{xie2023pixiu} & \textbf{0.94} & 0.85 & LLM \\
    \textit{FinMA-30B}~\cite{xie2023pixiu} & 0.88 & 0.87 & LLM \\
    \textit{FLANG-ELECTRA}~\cite{shah2022flue} & 0.92 & \textbf{0.92} & PLM \\
    \bottomrule
  \end{tabular}
  \label{tab:fin_benchmark}
\end{table}   

FinPLMs and FinLLMs fine-tuned on financial texts perform well on the FSA benchmark, such as \textit{PhraseBank (FPB)}~\cite{malo2014good} and \textit{FiQA-SA}~\cite{maia201818} datasets (results in Table~\ref{tab:fin_benchmark}). We observe that as a discriminative task, discriminative PLMs are more suitable than generative LLMs in FSA. Notably, the 0.11B-parameter PLM \textit{FLANG-ELECTRA}~\cite{shah2022flue} achieved top results: 92\% F1-score on \textit{FiQA-SA}~\cite{lee2025large} and 92\% on \textit{FPB} (only 2\% lower than \textit{FinMA-7B}), outperforming 5-shot generative LLMs and other PLMs.

Recent studies have examined sentiment links among related firms within the same industry. Jochem~\cite{jochem2019bias} found that bias propagates along supply chains, and Cao~\cite{cao2025too} reported that industry-level news can diffuse to firm-level sentiment. However, these studies do not explicitly model industry-specific sentiment.

Another key theme focuses on the temporal behavior of sentiment. Empirical evidence suggests that market reactions are delayed rather than immediate~\cite{smales2016news}. DeFond~\cite{defond2014timeliness} documents that sentiment effects fade over time. Despite this evidence, existing quantitative forecasting models that incorporate sentiment do not explicitly account for such lag effects.

\subsection{Bond Default Risk Forecasting}
\label{sec:related_works_bdrf}

From the perspectives of variable systems and feature sources, BDRF modeling has evolved from static cross-sectional settings to dynamic, multi-source frameworks. Early studies relied on cross-sectional firm-level financial indices and bond characteristics~\cite{Wu2023BondMutualFunds}, but such approaches ignore market transmission effects. With the development of short-term time series forecasting, dynamic macroeconomic indices have been incorporated to capture temporal dependence~\cite{JIANG2023}. Beyond structured variables, sentiment derived from unstructured text has also been introduced as an important feature source. Prior work shows that news sentiment affects asset pricing~\cite{aleti2024news} and improves forecasting performance in bond markets~\cite{erlwein2018macroeconomic,consoli2021emotions}. However, existing BDRF methods still capture a single firm-centric sentiment, so sentiment modeling for BDRF needs to shift from a firm-centric view to a multi-layered risk identification approach.
\begin{figure*}[t]
\centering
\includegraphics[width=0.75\textwidth]{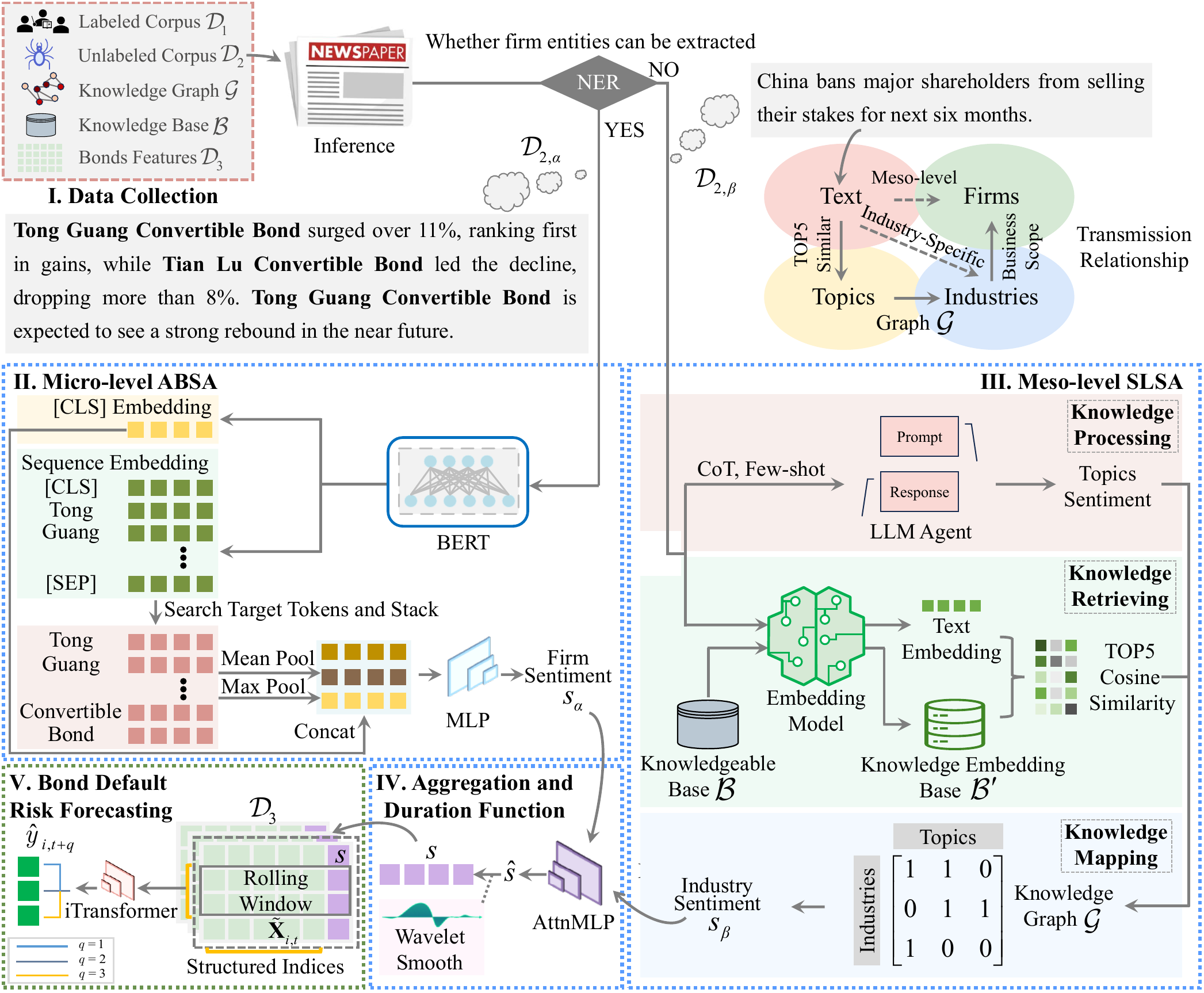}
\caption{\textbf{MulFSA Framework}. \uppercase\expandafter{\romannumeral1} (within the \textcolor[HTML]{D9958F}{red} box) is Task 1 Data Collection. \uppercase\expandafter{\romannumeral2}, \uppercase\expandafter{\romannumeral3}, \uppercase\expandafter{\romannumeral4} (within the \textcolor[HTML]{589BFF}{blue} box) belong to Task 2 Multi-level Sentiment Analysis. \uppercase\expandafter{\romannumeral5} (within the \textcolor[HTML]{69924E}{green} box) is Task 3 Bond Default Risk Forecasting.}
\label{fig:framework}
\end{figure*}

\section{Methodology}

We define the sentiment $s$ as a continuous value:

\begin{equation}
\small
\label{eq1}
    s \in [-1,1].
\end{equation}

There are $3$ polarities: \(s=-1\) indicates deteriorating risk; \(s=0\) is neutral, meaning no clear market impact; \(s=1\) signals positive drivers. 

MulFSA can be decomposed into three tasks depicted in Figure~\ref{fig:framework}: Task 1 is data collection (\S\ref{sec:data_collection}). Task 2 integrates sentiment extracted from texts across multiple views (\S\ref{sec:sentiment method}). Task 3 quantitatively verifies the effectiveness of the sentiment (\S\ref{sec:prediction}). To clarify the slightly overloaded notation, we provide a complete Symbol Reference in Appendix~\ref{appd:symbol_ref}.

\subsection{Task 1. Data Collection}
\label{sec:data_collection}
\renewcommand{\thefootnote}{\arabic{footnote}}

Currently, no public Chinese datasets meet our requirements, so we have constructed them ourselves. We need five datasets in total: \textbf{\textit{1)}} A labeled sentiment corpus \(\mathcal{D}_{1}\) for ABSA fine-tuning. \textit{\textbf{2)}} A Knowledge Graph \(\mathcal{G}\) of topics and a Knowledge Base \(\mathcal{B}\) containing topic definitions to support RAG. \textbf{\textit{3)}} A large-scale unlabeled corpus \(\mathcal{D}_{2}\) for inference the sentiment times series of all bonds. \textbf{\textit{4)}} A dataset \(\mathcal{D}_{3}\) of bonds with structured features for BDRF Modeling.

\textbf{\textit{Labeled Sentiment Corpus \(\mathcal{D}_{1}\)}}. Predicting probabilities instead of direct sentiment polarity helps the model learn the reasoning process~\cite{trisna2022deep}.
We downloaded 6,881 formal Chinese texts from \href{https://www.resset.com/index/home/}{\textit{RESSET}}
and \href{https://www.wind.com.cn/portal/en/EDB/index.html}{\textit{Wind}}, and perform labeling by a two-stage procedure (details in Appendix~\ref{appd:label}). Sentiment polarities must satisfy:
\begin{equation}
\small
\begin{aligned}
\mathbf{p}_s &= [p_{\text{neg}},\, p_{\text{neu}},\, p_{\text{pos}}], \\
\text{s.t.}\quad & p_{\text{neg}} + p_{\text{neu}} + p_{\text{pos}} = 1 .
\end{aligned}
\end{equation}
where $p_{\text{neg}}$, $p_{\text{neu}}$ and $p_{\text{pos}}$ are the probabilities of negative, neutral and positive polarity, respectively. 

\textbf{\textit{Knowledge Graph \(\mathcal{G}\) And Knowledge Base \(\mathcal{B}\)}}.
We selected 28 primary industries and 12 significant secondary industries according to \href{https://www.swsresearch.com/institute_sw/home}{\textit{SWS RESEARCH}}. We defined 117 topics across various domains: 27 general economic, 11 domestic economic, 14 global economic, 12 domestic financial, 13 international relations, 6 natural disasters, 10 technological advancements, 16 social developments, 7 environmental protection, and 7 legal regulations.
\(\mathcal{B}\) is the definition of all 117 topics. \(\mathcal{G}\) is a predefined Boolean matrix whose rows represent 40 industries and columns represent 117 topics:
\begin{equation}
\label{eq:graph}
\small
    \mathcal{G}=
        \begin{pmatrix}
        g_{1,1} & g_{1,2} & \cdots & g_{1,117} \\
        g_{2,1} & g_{2,2} & \cdots & g_{2,117} \\
        \vdots & \vdots & \ddots & \vdots \\
        g_{40,1} & g_{40,2} & \cdots & g_{40,117}
        \end{pmatrix}_{\substack{40 \times 117}}
        ,
\end{equation}
\begin{equation*}
    s.t. \quad g_{m,n}=\mathbbm{1}_{n \sim m}
\end{equation*}
where \(n \sim m\) denotes that topic \(n\) influences industry \(m\), with \(g_{m,n} = 1\) if the influence exists and $0$ otherwise. We provide an example subset of \(\mathcal{G}\) and \(\mathcal{B}\) in Appendix~\ref{appd:knowledge_graph}.

\textbf{\textit{Large-scale Unlabeled Corpus \(\mathcal{D}_{2}\)}.} 
Considering that user sentiment on social media exhibits significant fluctuations, which can introduce noise \cite{yue2019survey}, we selected formal texts concerning firms in the mainland China bond market as the research focus. We constructed a corpus with daily frequency spanning eleven years (2013 to 2023) from \href{http://hk.infobank.cn/}{\textit{Infobank}} and \href{https://www.wisers.com/}{\textit{WiseSearch}}, containing a total of 1,346,704 entries, with a daily frequency making the sentiment a medium-frequency factor. 

\textbf{\textit{Firms Features Dataset \(\mathcal{D}_{3}\)}.} 
We collected $6,472$ bonds from \href{https://ft.10jqka.com.cn/}{\textit{iFinD}} spanning from 2013 to 2023 with $45$ structured indices (the feature set taxonomy is in Appendix~\ref{appd:firms_features}). This dataset consists of three types of bonds: 47 below-A rated bonds, 407 defaulted bonds, and 6{,}018 matured corporate bonds. It is a time-series tabular dataset of daily frequency. The dataset was split for Task 3 at the bond level into training, validation, and test subsets in a 7:1:2 ratio.

We further divide \(\mathcal{D}_{2}\) into two subsets based on named entity recognition (NER) (see Appendix~\ref{appd:ner}): \(\mathcal{D}_{2,\alpha}\) containing a total of $623,444$ entries for micro-level ABSA inference and \(\mathcal{D}_{2,\beta}\) containing a total of $723,260$ entries for the meso-level SLSA inference. 
The statistical descriptions of datasets are in Figure~\ref{fig:dist}, and their entry structures are in Appendix~\ref{appd:stats}. The skewness of \(\mathcal{D}_{1}\), \(\mathcal{D}_{2,\alpha}\), and \(\mathcal{D}_{2,\beta}\) are $-0.066$, $-0.367$, and $5.358$. The long-tail effect of \(\mathcal{D}_{2,\beta}\) results from splitting, as it includes long announcements or reports (over $10,000$ tokens).

\subsection{Task 2. Multi-level Sentiment Extraction}
\label{sec:sentiment method}

Task 2 can be decomposed as follows: \textbf{\textit{1)}} \textbf{Micro-level ABSA} for firm-specific sentiment \(s_{\alpha}\): We fine-tuned a BERT model to perform ABSA on multiple firms mentioned in the same text. \textbf{\textit{2)}} \textbf{Meso-level SLSA} for industry-specific sentiment \(s_{\beta}\): We employed RAG based on knowledge graph and prompt engineering to interact with an LLM agent for SLSA. \textbf{\textit{3)}} \textbf{Aggregation} and \textbf{Duration Function}: Aggregation merges \(s_{\alpha}\) and \(s_{\beta}\) into the composite sentiment \(\hat{s}\), and the duration function \(h(\cdot)\) accounts for the effective duration of texts.

\begin{figure}[t]
    \centering
    \includegraphics[width=1.0\linewidth]{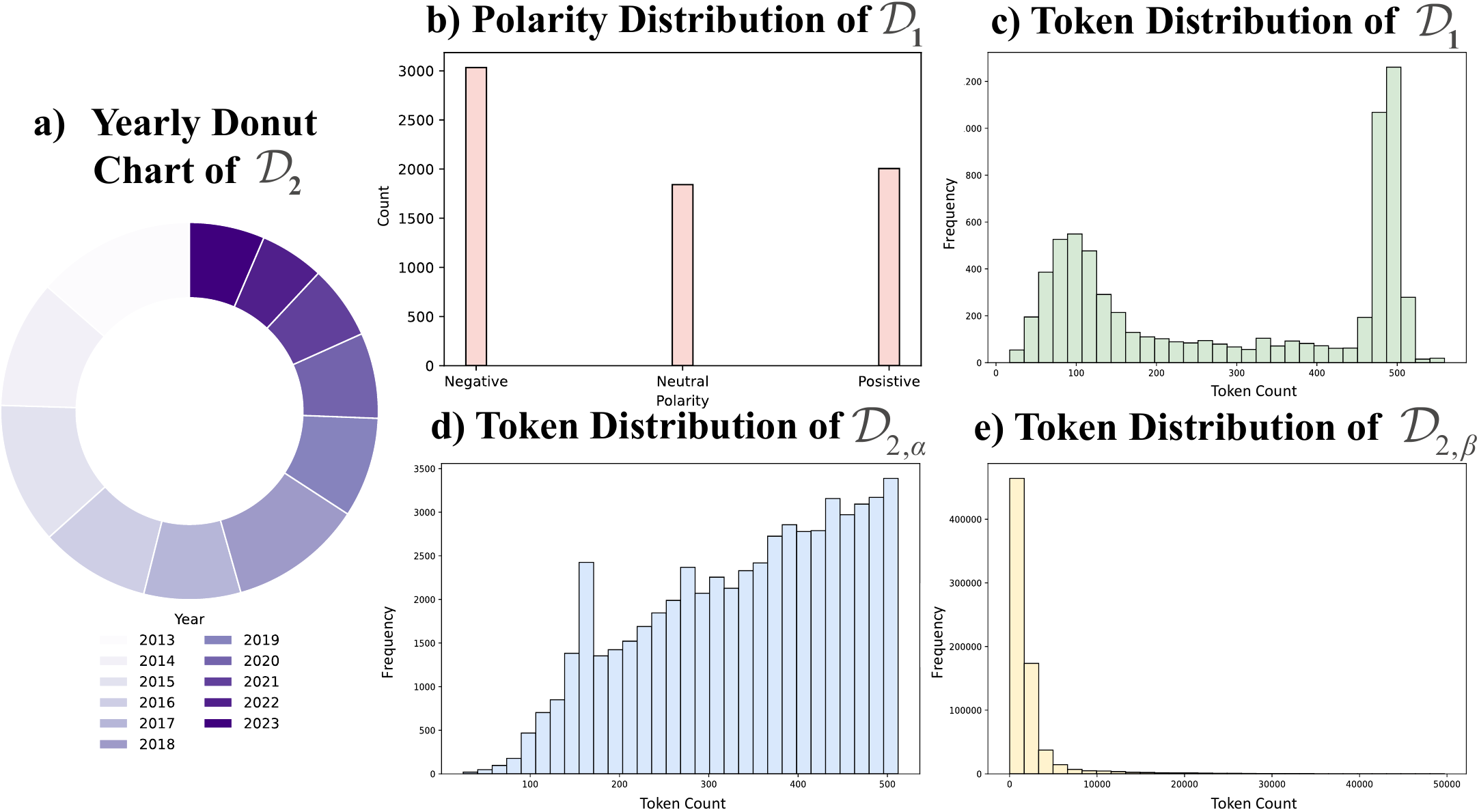}
    \caption{a) shows a donut chart of annual data volumes in \(\mathcal{D}_{2}\); b) shows the counts of three polarities in \(\mathcal{D}_{1}\); c), d), e) show token count distributions of \(\mathcal{D}_{1}\), \(\mathcal{D}_{2,\alpha}\), \(\mathcal{D}_{2,\beta}\).}
    \label{fig:dist}
\end{figure}

\textbf{\textit{Micro-level ABSA.}} This subtask is trained on labeled \(\mathcal{D}_{1}\) and inferred on unlabeled \(\mathcal{D}_{2,\alpha}\). As mentioned in \S\ref{sec:related_works_sa}, discriminative PLMs are more suitable than generative LLMs in FSA. Therefore, to be conducive to the more fine-grained ABSA, we fine-tuned a BERT followed by an MLP. Instead of treating ABSA as a sequence labeling problem, we directly output the sentiment for each entity in an End2End manner~\cite{min2023recent}, whose process of a single text is shown in the part \uppercase\expandafter{\romannumeral2} of Figure~\ref{fig:framework}. After encoding the text with BERT, we obtain sentence-level and token-level embeddings. For texts that mention multiple bonds, we select the embeddings of bond-related tokens and apply mean-max pooling \cite{zhang2018learning} to form a bond-level representation. This representation is concatenated with the sentence-level embedding and fed into an MLP to predict the sentiment polarity (pseudocode in Appendix~\ref{algo:mean-max-pool}). During inference, since \(\mathcal{D}_{2,\alpha}\) is at a daily frequency, for a given bond, there will either be excessive texts in a single day or none at all. Therefore, we denote as follows:
\begin{equation}
\label{eq2}
\small
    s_{\alpha,i,k}=\left\{
        \begin{aligned}
        & \frac{1}{J_{i}}\sum_{j=1}^{J_{i}} s_{\alpha,i,j,k}, J_{i} \geq 1  \\
        & 0, J_{i}=0 \\
        \end{aligned}
    \right.
    \quad ,
\end{equation}
where \(i\) is an arbitrary bond, \(j\) is an arbitrary text on a given day, \(J_{i}\) is the number of texts for a given bond, \(k\) is an arbitrary day, \(s_{\alpha,i,j,k}\) is the micro-level sentiment of a particular text for a specific bond on a given day calculated by the algorithm in Appendix~\ref{algo:mean-max-pool}. This equation implies that \(s_{\alpha,i,k}\) is the average of all sentiment values involving bond \(i\) on that day, thus preventing the accumulation of text volume from leading to extremes. After ABSA, we get the firm-specific sentiment matrix \(\mathcal{S_{\alpha}}\): 
\begin{equation}
\small
    \mathcal{S_{\alpha}}=
        \begin{pmatrix}
        s_{\alpha,1,1} & s_{\alpha,1,2} & \cdots & s_{\alpha,1,K} \\
        s_{\alpha,2,1} & s_{\alpha,2,2} & \cdots & s_{\alpha,2,K} \\
        \vdots & \vdots & \ddots & \vdots \\
        s_{\alpha,I,1} & s_{\alpha,I,2} & \cdots & s_{\alpha,I,K}
        \end{pmatrix}_{\substack{I \times K}}
        ,
\end{equation} 
where \(I\) denotes the number of firms and \(K\) the days in the time series. Since \(s_{\alpha,i,k}\) is an average within \([-1,1]\), \(s_{\alpha,i,k}\) also ranges in \([-1,1]\).


\textbf{\textit{Meso-level SLSA.}} The dataset used for SLSA is \(\mathcal{D}_{2,\beta}\). Since the relationship between firms and industries requires external knowledge, we regard an LLM as the agent for broader meso-level analysis, whose process of a single text is illustrated in the part \uppercase\expandafter{\romannumeral3} of Figure~\ref{fig:framework} consisting of three steps:
\begin{enumerate}
    \item Knowledge Processing. Each text passed through an LLM agent to obtain a sentiment polarity with few-shot learning and chain-of-thought (CoT) (prompt in Appendix~\ref{sec:prompt}). 
    \item Knowledge Retrieving. The embedding of text encoded by an embedding model compared with the topic embeddings in \(\mathcal{B}^{\prime}\) stored in the vector database by the cosine similarity. The top $5$ most similar topics are recalled, and the predicted sentiment is assigned to them. 
    \item Knowledge Mapping. \(\mathcal{G}\) is used as a bridge to map topics to downstream industries, thereby deriving industry-specific sentiment \(s_{\beta}\).
\end{enumerate}

Their pseudocode is in Appendix~\ref{algo:rag}. In \(\mathcal{D}_{3}\), we include the industry classification of each bond-issuing parent firm, and further match bonds to industries to obtain meso-level sentiment. During inference, similar to ABSA, we denote as follows:
\begin{equation}
\label{eq:meso} 
    \small
    \begin{aligned}
    s_{\beta,i,k} &= \frac{1}{M^{\ast}} \sum_{m=1}^{M^{\ast}} s_{\beta,m,k} \\
                  &= \left\{
                        \begin{aligned}
                             & \frac{1}{M^{\ast}} \sum_{m=1}^{M^{\ast}} \sum_{n=1}^{N^{\ast}} \sum_{j=1}^{J_{i}} c_{j,n} \cdot s_{\beta,n,j,k} \cdot g_{m,n}, J_{i} \geq 1\\
                             & 0, J_{i}=0 \\
                        \end{aligned}
                \right.
    \end{aligned}
    \quad ,
\end{equation}
where \(i\) is an arbitrary bond, \(M^{\ast}\) is the relevant industry to bond \(i\) , \(j\) is an arbitrary text on a given day, \(k\) is an arbitrary day, \(s_{\beta,n,j,k}\) is the sentiment of a particular text for a specific topic on a given day, \(J_{i}\) is the number of text for a given bond. \(g_{m,n}\) is the influence of topic \(n\) on industry \(m\) in Eq.~\ref{eq:graph}. \(N^{\ast}\) is the top $5$ most similar topics retrieved through \(\mathcal{B}^{\prime}\), with \(c_{j,n}\) denoting the 
cosine similarity between the embedding of text \(j\) and the embeddings of the top $5$ most similar topics \(N^{\ast}\). The transformation among topics, industries, and firms is illustrated in the subplot above the part \uppercase\expandafter{\romannumeral3} of Figure~\ref{fig:framework}:
\begin{itemize}
    \item \(g_{m,n}\) from \(\mathcal{G}\) enables the association between topics and industries, making the influence of industries on entities traceable.
    \item For macro events that affect all industries, $c_{j,n}$ controls the influence strength across different industries and bonds, allowing heterogeneous exposure across sectors.
    \item Beyond systemic events, industry-specific shocks are mapped to affected sectors via $\mathcal{G}$.
\end{itemize}
Regarding the absence of macro-level sentiment, macro factors such as monetary policy affect industries differently. Therefore, we decompose topic-level macro sentiment through \(\mathcal{G}\) and $c_{j,n}$, which is more accurate than applying a single macro-level sentiment to all bonds. After SLSA, we get the industry-specific sentiment matrix \(\mathcal{S_{\beta}}\): 
\begin{equation}
\small
    \mathcal{S_{\beta}}=
        \begin{pmatrix}
        s_{\beta,1,1} & s_{\beta,1,2} & \cdots & s_{\beta,1,K} \\
        s_{\beta,2,1} & s_{\beta,2,2} & \cdots & s_{\beta,2,K} \\
        \vdots & \vdots & \ddots & \vdots \\
        s_{\beta,M,1} & s_{\beta,M,2} & \cdots & s_{\beta,M,K}
        \end{pmatrix}_{\substack{M \times K}}
        ,
\end{equation}
where \(M\) is the number of industries and \(K\) is the number of days in the time series. \textit{z-score} standardization is applied to \(\mathcal{S_{\beta}}\) to mitigate sentiment extremes caused by excessive text volume and reduce discrepancies in \(s_{\beta,m,k}\). Since \(s_{\beta,i,k}\) is the average of multiple industry sentiments which may exceed $[-1,1]$, its range may also exceed $[-1,1]$.  

\textit{\textbf{Aggregation} and \textbf{Duration Function}}. We aggregate two types of sentiment by using an MLP:
\begin{equation}
    \small
    \hat{s}_{i,k}
    =
    \operatorname{AttnMLP}\!\left(
    \hat{s}_{\alpha,i,k},
    \hat{s}_{\beta,i,k}
    \right),
    \quad k = 1,\dots,K .
\end{equation}
where $\hat{s}_{i,k}$ denotes the aggregated sentiment time series of bond $i$ at time step $k$. When \( k = K \) is the current time, it uses historical data for real-time inference, and when \( k = K \) is a past date, it backtests. On days with fewer texts, a bond's sentiment is driven by its industry, reflecting lower market attention and heavier reliance on industry factors. The aggregation AttnMLP takes two sentiment series as input, maps each through a linear layer to latent embeddings, applies cross-attention between the embeddings, and then uses another linear layer with one-dimensional output to produce a sequence of the same length as the input.

However, not every bond or industry has daily text coverage, resulting in sparse time series with many neutral zeros. To address it, we design a duration function \(h(\cdot)\) to smooth \(\hat{s}_{i,k}\):
\begin{equation}
\small
    s_{i,k} = h(\hat{s}_{i,k}),
\end{equation}
where we choose \textit{wavelet smoothing} as the duration function \(h(\cdot)\), which has been proven effective in missing data imputation, and capturing data trends.

Notably, sentiment $s$ is initially continuous. In ABSA and SLSA, for convenience, it is converted into a polarity classification task. After aggregation and \(h(\cdot)\), the resulting values do not correspond to extreme points in most cases, and they may exceed the value range \([-1, 1]\) of individual texts.  

\newcommand{\X}{\mathbf{X}}
\newcommand{\y}{\mathbf{y}}

\subsection{Task 3. Bond Default Risk Forecasting}
\label{sec:prediction}

The dataset used for BDRF is \(\mathcal{D}_{3}\). We adopt a rolling window mechanism to treat this task as a time series forecasting problem, whose visual process is shown in Part \uppercase\expandafter{\romannumeral5} of Figure~\ref{fig:framework} and pseudocode is provided in Appendix~\ref{algo:bdp}. We take the features in \(\mathcal{D}_{3}\) and our composite sentiment \(s\) to forecast the proxy dependent variable \textit{credit spread} for default risk on the \(q\)-th day after the observation window \cite{liu2022novel,faust2013credit}, and employ A/B testing (with or without sentiment) to quantify the improvement of our extracted sentiment on downstream tasks such as BDRF:
\begin{equation}
\small
    \text{\textit{Credit Spread}} = \text{\textit{Bond Yield}} - \text{\textit{Risk-free Rate}},
\end{equation}
Since our main contribution is the sentiment extraction framework rather than the forecasting model, we use \textit{credit spread} forecasting as the backtesting task to validate the extracted sentiment only, instead of credit rating classification or other credit risk assessment tasks. The BDRF model is trained jointly with the aggregation AttnMLP in Task~2.

\begin{figure*}[t]
\centering
\includegraphics[width=0.85\textwidth]{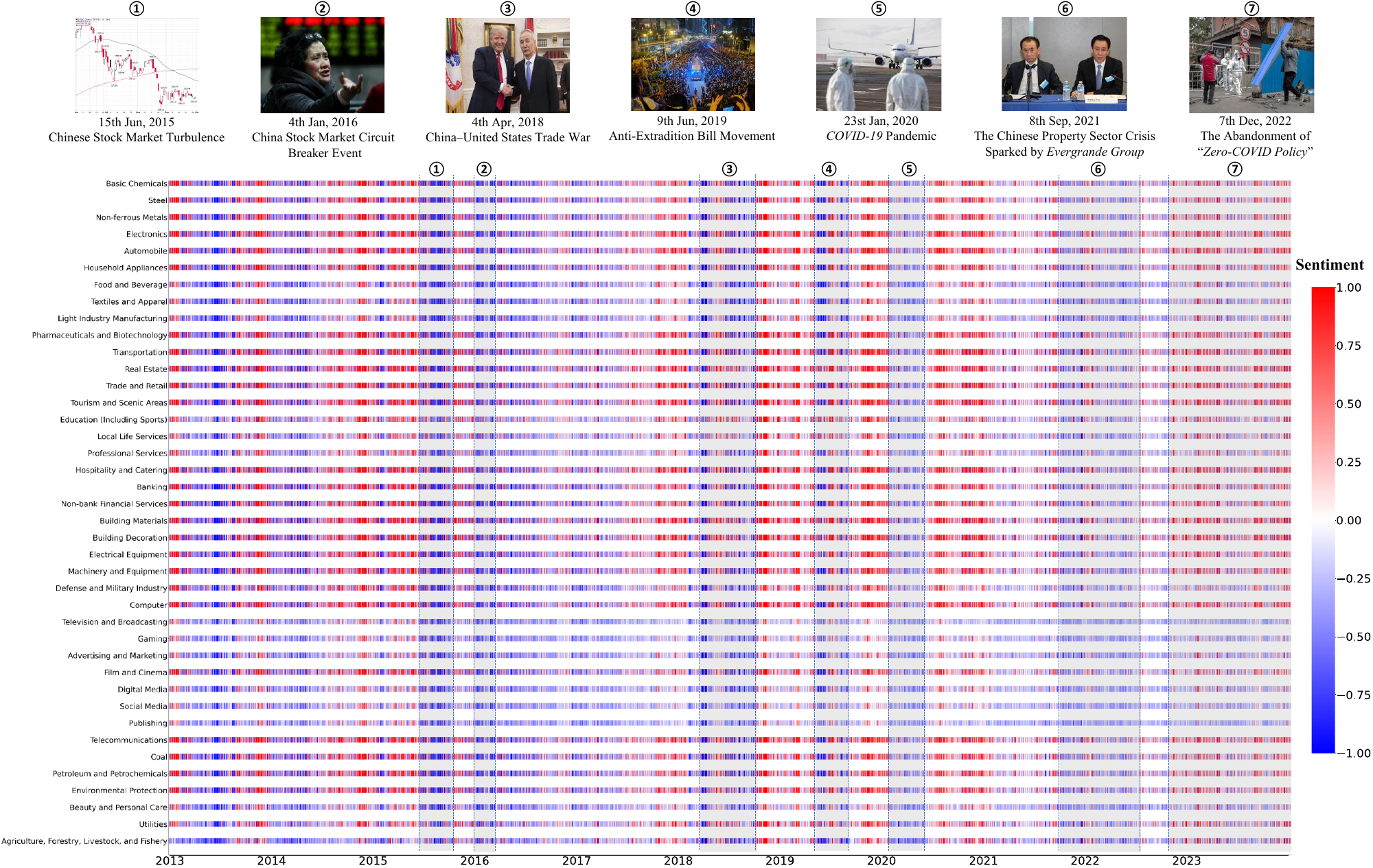}
\caption{\textbf{Sentiment Heatmap for 40 Industries}. Values exceeding $[-1,1]$ were truncated for display. The marked period exhibits a sentiment shift vis-à-vis its preceding period, aligning with the corresponding social event.}
\label{fig:year_line}
\end{figure*}

\section{Experiments}
\subsection{Experiments on MulFSA Components}
\label{sec:exp_setting}
\textit{\textbf{(i)}} \textbf{\textit{Model Selection.}} 
For ABSA BERT $f_{1}$, we used \textit{BERT-base-Chinese} and compared different global feature extraction methods, i.e., mean pooling and max pooling with our mean-max pooling.  
For the LLM Agent $f_{2}$, given that many FinLLMs are either closed-source (e.g., \textit{Agentar-Fin-R1}~\cite{zheng2025agentar}) or lack Chinese support (\textit{FinLLaVA}~\cite{huang2024open}, \textit{FinMA}~\cite{xie2023pixiu}), we used \textit{Qwen2.5-3B-Instruct} \cite{qwen2.5} as the foundation model and compared it with \textit{XuanYuan-6B}~\cite{zhang2023xuanyuan20largechinese} and other general LLMs under identical prompt settings.  
For Embedding Model $f_{3}$, we used \textit{bge-large-zh-v1.5} \cite{bge_embedding} to encode texts.  
For Forecasting Model $f_{4}$, we regressed \textit{credit spread} using iTransformer~\cite{liu2023itransformer} to validate the effectiveness of the sentiment, with the length of window $T=21$.  
We chose the \textit{Daubechies 4 wavelet} as the basis function for $h(\cdot)$, with the level set to $6$.
\textit{\textbf{(ii)}} \textbf{\textit{Metrics.}} 
We used Precision to measure the output of $f_{1}$ and $f_{2}$, MAE and MAPE to measure the \textit{credit spread} output of $f_{4}$ (calculation in Appendix~\ref{appd:metrics}).
\textit{\textbf{(iii)}} \textbf{\textit{Implementation Details.}} 
Hyperparameters and training details are in Appendix~\ref{appd:setting}.

Experimental results in Appendix~\ref{appd:ner} show our NER component has a Precision of $94.55\%$, and results in Appendix~\ref{exp:micro} and Appendix~\ref{exp:meso} show that our model selection choice is reasonable, where ABSA BERT $f_{1}$ for micro-level FSA could achieve the Precision of $88.27\%$ and LLM Agent $f_{2}$ for meso-level ABSA could achieve the Precision of $75.0\%$. The \textit{Deepseek-R1} technical report~\cite{deepseekai2025deepseekr1incentivizingreasoningcapability} notes its prompt sensitivity, with few-shot prompting consistently degrading performance. An interesting observation is that \textit{Qwen2.5-3B-Instruct} basically outperforms other models in meso-level SLSA. Therefore, we attribute this to the fact that our prompts adopt CoT and few-shot learning, and \textit{Qwen2.5-3B-Instruct}, as a model with high knowledge density, is more suitable for this task.

\subsection{Visualization of Industry Sentiment}

We visualized the industry-specific sentiment matrix \(\mathcal{S_{\beta}}\) in Figure~\ref{fig:year_line}. Clearly, we observe the following: 
\textit{\textbf{(i)}} On a specific day or short period, sentiment across industries may vary.
\textit{\textbf{(ii)}} In the long run, all industries exhibit similar sentiment patterns, with changes following the same overall trend, i.e., collective optimism or pessimism. It endorses \textit{the financial contagion} \cite{forbes2002no} and \textit{the peer effect} \cite{falk2006clean}, which illustrates how economic shocks or significant events in one market can lead to simultaneous reactions spreading across multiple markets. At certain time points, all industries experience collective sentiment shifts. Such events act as a ``curtain" for most industries, instantly impacting sector-wide sentiment and persisting over time. We provide a more detailed explanation of Figure~\ref{fig:year_line} in Appendix~\ref{appd:exp on fig3}.

\begin{figure}[!t]
    \centering
    \includegraphics[width=0.8\linewidth]{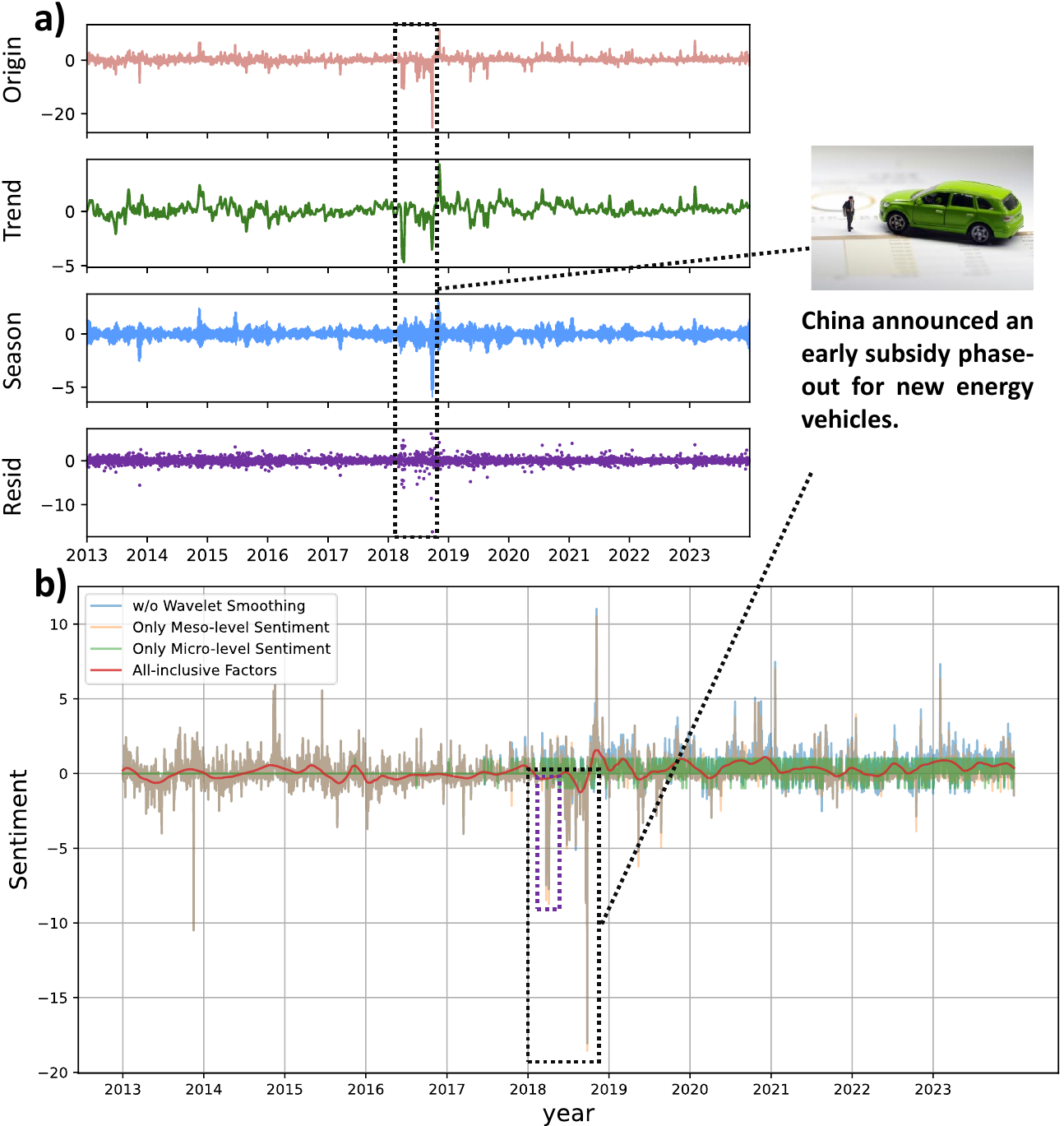}
    \caption{\textbf{a) Industry Sentiment Time Series Decomposition of \textit{Automobile} (2013-2023)}. From top to bottom, the four subplots represent the origin, trend, seasonality, and residuals. \textbf{b) Component Visualization of \textit{CATL}}. Since industry sentiment is derived from topic sentiment, which is aggregated from daily sentiment, identical polarities across all related industries on certain days can produce extreme values in the visualization.}
    \label{fig:decomposition}
\end{figure}

\subsection{Empirical Analysis on BDRF Modeling}
\label{exp:BDRF}

To quantitatively validate our composite sentiment of each bond, we selected BDRF as the downstream task, taking \textit{credit spread} as the dependent variable.

First, to show how \textit{credit spread} forecasting performance varies across different targets \(q\) when incorporating our extracted sentiment, we conducted comparative experiments. Results are in Table~\ref{tab:exp_targets}, where (\CheckmarkBold) indicates adding composite sentiment into the feature set, with the $p$-value calculated by a non-parametric \textit{Permutation Test} (50 resamples, 95\% confidence interval)~\cite{ojala2010permutation}. \(\Delta\)MAE and \(\Delta\)MAPE in the table denote the percentage change in MAE and MAPE with sentiment, computed by subtracting the corresponding metric of the with-sentiment group from that of the without-sentiment group. That is, a positive value indicates an increase, while a negative value indicates a decrease. We found that: \textit{\textbf{(i)}} As $q$ increases, forecasting errors grow with or without sentiment, reflecting the difficulty of forecasting longer-horizon default risk under limited information and high-frequency price fluctuations. \textit{\textbf{(ii)}} Sentiment from unstructured text captures market mood and acts as an exogenous feature that improves forecasting performance. The largest gain appears at $t+4$, which we attribute to the latency and persistence of textual impact, as sentiment diffuses gradually and becomes more distinguishable from noise at medium-term horizons. Since the settlement cycle for bonds in the Chinese securities market is $t+1$, the forecast horizon in subsequent BDRF experiments is set to $t+1$ with $T = 21$.

\begin{table}[!t]
    \centering
    \captionsetup{width=\linewidth}
    \caption{Comparison of BDRF Results across Forecasting Targets $q$ (w/ vs. w/o sentiment).}
    \label{tab:exp_targets}
    \tiny
    \resizebox{\linewidth}{!}{ 
        \setlength\tabcolsep{3pt} 
        \begin{tabular}{cccccccc} 
        \toprule
        \multicolumn{2}{c}{\makecell{Target\\$t+q$}} & Sent. 
        & \makecell{MAE\\(e-5)} & \makecell{MAPE\\(e-3)} & $p$ 
        & \makecell{$\Delta$MAE\\(\%$\downarrow$)} & \makecell{$\Delta$MAPE\\(\%$\downarrow$)} \\
        \midrule
        \multicolumn{2}{c}{$t+1$} & & 8.39 & 8.29 & n/a & n/a & n/a \\ 
        \multicolumn{2}{c}{$t+1$} & \CheckmarkBold & 7.53 & 7.30 & 0.041 & 10.25 & 11.94 \\ 
        \multicolumn{2}{c}{$t+2$} & & 8.31 & 8.76 & n/a & n/a & n/a \\ 
        \multicolumn{2}{c}{$t+2$} & \CheckmarkBold & 7.98 & 7.74 & 0.035 & 3.97 & 11.64 \\ 
        \multicolumn{2}{c}{$t+3$} & & 11.13 & 9.32 & n/a & n/a & n/a \\ 
        \multicolumn{2}{c}{$t+3$} & \CheckmarkBold & 10.37 & 8.12 & 0.063 & 6.82 & 12.87 \\ 
        \multicolumn{2}{c}{$t+4$} & & 13.29 & 8.74 & n/a & n/a & n/a \\ 
        \multicolumn{2}{c}{$t+4$} & \CheckmarkBold & 11.27 & 7.23 & 0.036 & 15.19 & 17.27 \\ 
        \bottomrule
        \end{tabular}
    }
\end{table}

\begin{table}[!t]
    \centering
    \captionsetup{width=\linewidth}
    \caption{Comparison of Baselines vs. Ours.}
    \label{tab:exp_baselines}
    \resizebox{\linewidth}{!}{ 
        \setlength\tabcolsep{6pt}
        \begin{tabular}{lcc} 
        \toprule
        Baseline Model & \makecell{MAE\\(e-5)} & \makecell{MAPE\\(e-3)} \\
        \midrule
        \multicolumn{3}{c}{\cellcolor{yellow!40}\makebox[\linewidth]{\makecell{\textit{Different Sentiment Analysis Models under Our \(f_{4}\)}}}}\\
        \textit{Baichuan-7B}~\cite{yang2023baichuan} & 14.29 & 14.02 \\
        \textit{Baichuan-7B w/ wavelet} & 13.87 & 13.50 \\
        \textit{DeepSeek-R1-Distill-Qwen-7B}~\cite{deepseekai2025deepseekr1incentivizingreasoningcapability} & 13.29 & 12.48 \\
        \textit{DeepSeek-R1-Distill-Qwen-7B w/ wavelet} & 13.04 & 12.17 \\
        \textit{XuanYuan-6B}~\cite{zhang2023xuanyuan20largechinese} & 11.80 & 10.43 \\
        \textit{XuanYuan-6B w/ wavelet} & 11.27 & 9.93 \\
        \multicolumn{3}{c}{\cellcolor{yellow!40}\makebox[\linewidth]{\makecell{\textit{Different Forecasting Models with Our Sentiment}}}}\\
        \textit{Random Forest} & 55.82 & 30.28 \\
        \textit{XGBoost} & 23.49 & 17.21 \\
        \textit{LSTM} & 18.48 & 13.58 \\
        \textit{Informer}~\cite{zhou2021informer} & 11.37 & 11.21 \\
        \textit{Chronos-2 (zero-shot)}~\cite{ansari2025chronos2} & 9.22 & 9.07 \\
        \textit{PatchTST}~\cite{Yuqietal-2023-PatchTST} & 8.72 & 8.48 \\
        \midrule
        \textbf{Our \(f^{*}\)} & \textbf{7.53} & \textbf{7.30} \\
        \bottomrule
        \end{tabular}
    }
\end{table}

\begin{table}[!t]
    \centering
    \captionsetup{width=\linewidth}
    \caption{Ablation Study on MulFSA Components.}
    \label{tab:exp_ablation}
    \tiny
    \resizebox{\linewidth}{!}{
        \setlength\tabcolsep{2pt}
        \begin{tabular}{cccccccc} 
        \toprule
        \makecell{Micro\\Sent.} & \makecell{Meso\\Sent.} & \makecell{Dur.\\Func.} 
        & \makecell{MAE\\(e-5)} & \makecell{MAPE\\(e-3)} & $p$ 
        & \makecell{$\Delta$MAE\\(\%$\downarrow$)} & \makecell{$\Delta$MAPE\\(\%$\downarrow$)} \\
        \midrule
         & & & 8.39 & 8.29 & n/a & n/a & n/a \\  
        \CheckmarkBold & & & 8.27 & 9.06 & 0.059 & 1.43 & -9.28 \\ 
         &\CheckmarkBold & & 8.19 & 8.41 & 0.042 & 2.38 & -1.44 \\ 
        \CheckmarkBold &\CheckmarkBold & & 13.58 & 41.47 & $\approx$ 0.0 & -61.85 & -400.24 \\
        * & * & & 8.42 & 8.18 & 0.096 & -0.35 & 1.32 \\
        \CheckmarkBold & \CheckmarkBold & \CheckmarkBold & \textbf{7.53} & \textbf{7.30} & \textbf{0.041} & \textbf{10.25} & \textbf{11.94} \\ 
        \bottomrule
        \end{tabular}
    }
\end{table}

Second, to compare our $f^{*}$ with other baselines, we conducted comparative experiments with results in Table~\ref{tab:exp_baselines}. The first $6$ rows show LLM-extracted sentiment results under our $f_{4}$. During inference, each LLM outputs a sentiment value for each text in $\mathcal{D}_{2,\beta}$ directly. To match the $f_4$ formulation, we aggregate sentiment by taking the daily mean. Due to the large scale of $\mathcal{D}_{2,\beta}$, the resulting sentiment series is relatively continuous, which weakens the effect of the Duration Function on baseline LLMs. The following $6$ rows show the results of different forecasting models with our sentiment. \textit{Random Forest} and \textit{XGBoost} directly forecast based on cross-sectional data, while \textit{LSTM}, \textit{Informer}, \textit{Chronos-2}, and \textit{PatchTST} use the same rolling window with $f_{4}$. Results indicate $f^{*}$ outperforms all baselines, being more suitable for FSA.

Third, to assess the three components in Task 2 under partial integration, we performed an ablation study with results in Table~\ref{tab:exp_ablation}, where the first row shows results without sentiment; (\CheckmarkBold) denotes included components; in the fifth row, (*) indicates micro- and meso-level sentiments as two separate features (not aggregation). Adding micro- or meso-level sentiment alone slightly improves forecasting performance (1.43\% and 2.38\% reductions in $\Delta$MAE). For the 6{,}472 bonds in $\mathcal{D}_{3}$, we compute the orthogonality between micro- and meso-level sentiment for each bond and report the average across bonds: Pearson = 0.0218, Spearman = 0.0321, and cosine similarity = 0.0283, indicating that the two components are approximately orthogonal. The addition of two individually useful features \(s_{\alpha}\) and \(s_{\beta}\) leads to a performance drop (a 400\% degradation) because market attention varies across firms: Some have abundant texts, while others have very few, making the micro-level sentiment of low-attention firms a sparse time series (sparse meaning the sentiment is nonzero on only a few days). Their occasional nonzero values cause abrupt fluctuations during aggregation, increasing randomness. This highlights the need for the duration function to filter such abnormal noise.

We visualized components using \textit{Contemporary Amperex Technology Co., Limited (CATL)} in Figure~\ref{fig:decomposition} b), a representative \textit{Automobile} industry supplier, obtaining two observations: \textit{\textbf{(i)}} Individual components without smoothing fluctuate sharply, poorly reflecting actual risk dynamics. However, after aggregation and the application of the duration function, the all-inclusive factors curve captures overall public sentiment trends effectively. \textit{\textbf{(ii)}} \textit{CATL} and the \textit{Automobile} industry show similar bearish trends in the marked period, but \textit{CATL} is more pessimistic on certain days. This indicates the company is affected not only by industry-wide bearish sentiment transmission but also by direct targeting via numerous bearish market texts. In Appendix~\ref{appd:duration}, we provide comparative experiments on different duration functions to demonstrate the effectiveness of \textit{wavelet smoothing}.

Finally, to examine the collinearity among features in \(f^{*}\) and their contributions during forecasting, we conducted a feature analysis. We plotted a correlation matrix in Appendix~\ref{appd:corr mat}, which shows that our constructed sentiment does not exhibit significant collinearity with other features. We presented the feature attribution of $f^{*}$ in Appendix~\ref{appd:feature_importance}, where the feature importance is calculated by randomly permuting the positions of independent variables~\cite{altmann2010permutation}. We observe that in the forecasting of short-term \textit{credit spread}, \textit{Macroeconomic and Financial Indicators} contribute the most. Our composite sentiment ranks $9$th among the $46$ features, comparable to some highly contributing firm-level indicators such as \textit{Manufacturing PMI}. In Appendix~\ref{appd:case}, we present visualizations of sentiment and their corresponding \textit{credit spreads} for several bonds before default, together with case studies.

\section*{Limitations}
\label{sec:conclusion}
We acknowledge the following limitations:
\begin{enumerate}
    \item First, our work is essentially an FSA paradigm for modeling firm entities, which can measure the risk premia of all securities issued by a given entity. However, it has not yet been extended to other countries, nor to other asset classes and downstream tasks in the financial market, including equities and derivatives.
    \item Second, we use the Knowledge Graph $\mathcal{G}$ and the Knowledge Base $\mathcal{B}$. In real-world settings, the operating conditions of bond issuers’ parent companies and the impacts of different industry topics change over time. In contrast, $\mathcal{G}$ and $\mathcal{B}$ constructed in this work are static. This means that, for MulFSA to be deployed in an industrial setting, it would need the ability to automatically acquire updated data, which also implies automatically collecting relevant news about the target entities.
    \item Finally, as mentioned in \S\ref{sec:exp_setting}, due to the language support and open-source limitations, MulFSA relies on multiple models, including relatively old models such as BERT. If a future FinLLM supports Chinese and shows strong performance across financial benchmarks, a more promising solution would be to complete Task~2 multi-level sentiment analysis using a single unified model after supervised fine-tuning (SFT) to make it more adaptable to financial contexts.
\end{enumerate}

\section*{Acknowledgments}
We would like to thank the support from the National Natural Science Foundation of China under Grants No. 72371178.


\bibliography{main}

@article{JIANG2023,
author = {Fuwei, Jiang and others},
title = {Research on an Early Warning Model of Corporate Bond Default and its Economic Mechanism Based on Machine Learning},
publisher = {Journal of Financial Research},
year = {2023},
journal = {Journal of Financial Research},
volume = {520},
number = {10},
eid = {85},
numpages = {18},
pages = {85},
doi = {}
}

@article{araci2019finbert,
  title={Finbert: Financial sentiment analysis with pre-trained language models},
  author={Araci, Dogu},
  journal={arXiv preprint arXiv:1908.10063},
  year={2019}
}

@article{achiam2023gpt,
  title={GPT-4 technical report},
  author={OpenAI},
  journal={arXiv preprint arXiv:2303.08774},
  year={2023}
}

@article{Wu2023BondMutualFunds,
  author  = {Wu, Yuhui and Liu, Xiaoling and Qi, Shusen},
  title   = {Hurting the Innocent: The Role of Bond Mutual Funds in Propagating Idiosyncratic Bond Defaults},
  journal = {The Journal of World Economy},
  year    = {2023},
  volume  = {46},
  number  = {8},
  pages   = {186-210}
}

@article{wu2023bloomberggpt,
  title={Bloomberggpt: A large language model for finance},
  author={Wu, Shijie and Irsoy, Ozan and Lu, Steven and Dabravolski, Vadim and Dredze, Mark and Gehrmann, Sebastian and Kambadur, Prabhanjan and Rosenberg, David and Mann, Gideon},
  journal={arXiv preprint arXiv:2303.17564},
  year={2023}
}

@article{lee2025large,
  title={Large Language Models in Finance (FinLLMs)},
  author={Lee, Jean and Stevens, Nicholas and Han, Soyeon Caren},
  journal={Neural Computing and Applications},
  pages={1--15},
  year={2025},
  publisher={Springer}
}

@article{shah2022flue,
  title={When flue meets flang: Benchmarks and large pre-trained language model for financial domain},
  author={Shah, Raj Sanjay and Chawla, Kunal and Eidnani, Dheeraj and Shah, Agam and Du, Wendi and Chava, Sudheer and Raman, Natraj and Smiley, Charese and Chen, Jiaao and Yang, Diyi},
  journal={arXiv preprint arXiv:2211.00083},
  year={2022}
}

@article{malo2014good,
  title={Good debt or bad debt: Detecting semantic orientations in economic texts},
  author={Malo, Pekka and Sinha, Ankur and Korhonen, Pekka and Wallenius, Jyrki and Takala, Pyry},
  journal={Journal of the Association for Information Science and Technology},
  volume={65},
  number={4},
  pages={782--796},
  year={2014},
  publisher={Wiley Online Library}
}

@inproceedings{maia201818,
  title={Www'18 open challenge: financial opinion mining and question answering},
  author={Maia, Macedo and Handschuh, Siegfried and Freitas, Andr{\'e} and Davis, Brian and McDermott, Ross and Zarrouk, Manel and Balahur, Alexandra},
  booktitle={Companion proceedings of the the web conference 2018},
  pages={1941--1942},
  year={2018}
}

@article{zhang2018learning,
  title={Learning universal sentence representations with mean-max attention autoencoder},
  author={Zhang, Minghua and Wu, Yunfang and Li, Weikang and Li, Wei},
  journal={arXiv preprint arXiv:1809.06590},
  year={2018}
}

@article{faust2013credit,
  title={Credit spreads as predictors of real-time economic activity: a Bayesian model-averaging approach},
  author={Faust, Jon and Gilchrist, Simon and Wright, Jonathan H and Zakraj{\v{s}}sek, Egon},
  journal={Review of Economics and Statistics},
  volume={95},
  number={5},
  pages={1501--1519},
  year={2013},
  publisher={The MIT Press}
}

@misc{deepseekai2025deepseekr1incentivizingreasoningcapability,
      title={DeepSeek-R1: Incentivizing Reasoning Capability in LLMs via Reinforcement Learning}, 
      author={DeepSeek-AI},
      year={2025},
      eprint={2501.12948},
      archivePrefix={arXiv},
      primaryClass={cs.CL},
}

@misc{bge_embedding,
      title={C-Pack: Packaged Resources To Advance General Chinese Embedding}, 
      author={Shitao Xiao and Zheng Liu and Peitian Zhang and Niklas Muennighoff},
      year={2023},
      eprint={2309.07597},
      archivePrefix={arXiv},
      primaryClass={cs.CL}
}

@article{forbes2002no,
  title={No contagion, only interdependence: measuring stock market comovements},
  author={Forbes, Kristin J and Rigobon, Roberto},
  journal={The journal of Finance},
  volume={57},
  number={5},
  pages={2223--2261},
  year={2002},
  publisher={Wiley Online Library}
}

@article{du2024financial,
  title={Financial sentiment analysis: Techniques and applications},
  author={Du, Kelvin and Xing, Frank and Mao, Rui and Cambria, Erik},
  journal={ACM Computing Surveys},
  volume={56},
  number={9},
  pages={1--42},
  year={2024},
  publisher={ACM New York, NY}
}

@article{ojala2010permutation,
  title={Permutation tests for studying classifier performance.},
  author={Ojala, Markus and Garriga, Gemma C},
  journal={Journal of machine learning research},
  volume={11},
  number={6},
  year={2010}
}

@article{aleti2024news,
  title={News and asset pricing: A high-frequency anatomy of the sdf},
  author={Aleti, Saketh and Bollerslev, Tim},
  journal={The Review of Financial Studies},
  pages={hhae019},
  year={2024},
  publisher={Oxford University Press}
}

@article{yue2019survey,
  title={A survey of sentiment analysis in social media},
  author={Yue, Lin and others},
  journal={Knowledge and Information Systems},
  volume={60},
  pages={617--663},
  year={2019},
  publisher={Springer}
}

@article{erlwein2018macroeconomic,
  title={Macroeconomic news sentiment: enhanced risk assessment for sovereign bonds},
  author={Erlwein-Sayer, Christina},
  journal={Risks},
  volume={6},
  number={4},
  pages={141},
  year={2018},
  publisher={MDPI}
}

@article{consoli2021emotions,
  title={Emotions in macroeconomic news and their impact on the european bond market},
  author={Consoli, Sergio and Pezzoli, Luca Tiozzo and Tosetti, Elisa},
  journal={Journal of International Money and Finance},
  volume={118},
  pages={102472},
  year={2021},
  publisher={Elsevier}
}

@article{perri2018international,
  title={International recessions},
  author={Perri, Fabrizio and Quadrini, Vincenzo},
  journal={American Economic Review},
  volume={108},
  number={4-5},
  pages={935--984},
  year={2018},
  publisher={American Economic Association 2014 Broadway, Suite 305, Nashville, TN 37203}
}

@inproceedings{ahbali2022identifying,
  title={Identifying corporate credit risk sentiments from financial news},
  author={Ahbali and others},
  booktitle={Proceedings of the 2022 Conference of the North American Chapter of the Association for Computational Linguistics: Human Language Technologies: Industry Track},
  pages={362--370},
  year={2022}
}

@article{falk2006clean,
  title={Clean evidence on peer effects},
  author={Falk, Armin and Ichino, Andrea},
  journal={Journal of labor economics},
  volume={24},
  number={1},
  pages={39--57},
  year={2006},
  publisher={The University of Chicago Press}
}

@article{efretuei2021year,
  title={Year and industry-level accounting narrative analysis: Readability and tone variation},
  author={Efretuei, Ekaete},
  journal={Journal of Emerging Technologies in Accounting},
  volume={18},
  number={2},
  pages={53--76},
  year={2021},
  publisher={American Accounting Association}
}

@article{benhabib2016sentiments,
  title={Sentiments, financial markets, and macroeconomic fluctuations},
  author={Benhabib, Jess and Liu, Xuewen and Wang, Pengfei},
  journal={Journal of Financial Economics},
  volume={120},
  number={2},
  pages={420--443},
  year={2016},
  publisher={Elsevier}
}

@article{jochem2019bias,
  title={Bias propagation in economically linked firms},
  author={Jochem, Torsten and Peters, Florian S},
  journal={Available at SSRN 2698365},
  year={2019}
}

@article{cao2025too,
  title={Too sensitive to fail: The impact of sentiment connectedness on stock price crash risk},
  author={Cao, Jie and He, Guoqing and Jiao, Yaping},
  journal={Entropy},
  volume={27},
  number={4},
  pages={345},
  year={2025},
  publisher={MDPI}
}

@article{defond2014timeliness,
  title={The timeliness of the bond market reaction to bad earnings news},
  author={Defond, Mark L and Zhang, Jieying},
  journal={Contemporary Accounting Research},
  volume={31},
  number={3},
  pages={911--936},
  year={2014},
  publisher={Wiley Online Library}
}

@article{smales2016news,
  title={News sentiment and bank credit risk},
  author={Smales, Lee A},
  journal={Journal of Empirical Finance},
  volume={38},
  pages={37--61},
  year={2016},
  publisher={Elsevier}
}

@techreport{goetzmann2022crash,
  title={Crash narratives},
  author={Goetzmann, William N and Kim, Dasol and Shiller, Robert J},
  year={2022},
  institution={National Bureau of Economic Research}
}

@article{altmann2010permutation,
  title={Permutation importance: a corrected feature importance measure},
  author={Altmann, Andr{\'e} and Tolo{\c{s}}i, Laura and Sander, Oliver and Lengauer, Thomas},
  journal={Bioinformatics},
  volume={26},
  number={10},
  pages={1340--1347},
  year={2010},
  publisher={Oxford University Press}
}

@article{li2024extracting,
  title={Extracting financial data from unstructured sources: Leveraging large language models},
  author={Li, Huaxia and Gao, Haoyun and Wu, Chengzhang and Vasarhelyi, Miklos A},
  journal={Journal of Information Systems},
  pages={1--22},
  year={2024},
  publisher={American Accounting Association}
}

@article{min2023recent,
  title={Recent advances in natural language processing via large pre-trained language models: A survey},
  author={Min, Bonan and Ross, Hayley and Sulem, Elior and Veyseh, Amir Pouran Ben and Nguyen, Thien Huu and Sainz, Oscar and Agirre, Eneko and Heintz, Ilana and Roth, Dan},
  journal={ACM Computing Surveys},
  volume={56},
  number={2},
  pages={1--40},
  year={2023},
  publisher={ACM New York, NY}
}

@inproceedings{gong2020unified,
  title={Unified feature and instance based domain adaptation for aspect-based sentiment analysis},
  author={Gong, Chenggong and Yu, Jianfei and Xia, Rui},
  booktitle={Proceedings of the 2020 Conference on EMNLP},
  pages={7035--7045},
  year={2020}
}

@article{ehrmann2023named,
  title={Named entity recognition and classification in historical documents: A survey},
  author={Ehrmann, Maud and Hamdi, Ahmed and Pontes, Elvys Linhares and Romanello, Matteo and Doucet, Antoine},
  journal={ACM Computing Surveys},
  volume={56},
  number={2},
  pages={1--47},
  year={2023},
  publisher={ACM New York, NY}
}

@inproceedings{huang2020entity,
  title={An entity-level sentiment analysis of financial text based on pre-trained language model},
  author={Huang, Zhihong and Fang, Zhijian},
  booktitle={2020 IEEE 18th International Conference on Industrial Informatics (INDIN)},
  volume={1},
  pages={391--396},
  year={2020},
  organization={IEEE}
}

@article{trisna2022deep,
  title={Deep learning approach for aspect-based sentiment classification: a comparative review},
  author={Trisna, Komang Wahyu and Jie, Huang Jin},
  journal={Applied Artificial Intelligence},
  volume={36},
  number={1},
  pages={2014186},
  year={2022},
  publisher={Taylor \& Francis}
}

@article{zheng2025agentar,
  title={Agentar-Fin-R1: Enhancing Financial Intelligence through Domain Expertise, Training Efficiency, and Advanced Reasoning},
  author={Zheng, Yanjun and others},
  journal={arXiv preprint arXiv:2507.16802},
  year={2025}
}

@article{huang2024open,
  title={Open-finllms: Open multimodal large language models for financial applications},
  author={Huang, Jimin and Xiao, Mengxi and Li, Dong and Jiang, Zihao and Yang, Yuzhe and Zhang, Yifei and Qian, Lingfei and Wang, Yan and Peng, Xueqing and Ren, Yang and others},
  journal={arXiv preprint arXiv:2408.11878},
  year={2024}
}

@article{xie2023pixiu,
  title={Pixiu: A comprehensive benchmark, instruction dataset and large language model for finance},
  author={Xie, Qianqian and Han, Weiguang and Zhang, Xiao and Lai, Yanzhao and Peng, Min and Lopez-Lira, Alejandro and Huang, Jimin},
  journal={Advances in Neural Information Processing Systems},
  volume={36},
  pages={33469--33484},
  year={2023}
}

@article{huang2023finbert,
  title={FinBERT: A large language model for extracting information from financial text},
  author={Huang, Allen H and Wang, Hui and Yang, Yi},
  journal={Contemporary Accounting Research},
  volume={40},
  number={2},
  pages={806--841},
  year={2023},
  publisher={Wiley Online Library}
}

@article{llama3modelcard,

title={Llama 3 Model Card},

author={AI@Meta},

year={2024},

url = {https://github.com/meta-llama/llama3/tree/main}

}

@misc{zhang2023xuanyuan20largechinese,
      title={XuanYuan 2.0: A Large Chinese Financial Chat Model with Hundreds of Billions Parameters}, 
      author={Xuanyu Zhang and Qing Yang and Dongliang Xu},
      year={2023},
      eprint={2305.12002},
      archivePrefix={arXiv},
      primaryClass={cs.CL},
      url={https://arxiv.org/abs/2305.12002}, 
}

@article{yang2023baichuan,
  title={Baichuan 2: Open large-scale language models},
  author={Baichuan Inc.},
  journal={arXiv preprint arXiv:2309.10305},
  year={2023}
}

@misc{qwen2.5,
    title = {Qwen2.5: A Party of Foundation Models},
    url = {https://qwenlm.github.io/blog/qwen2.5/},
    author = {Team Qwen},
    month = {September},
    year = {2024}
}

@online{Company-Names-Corpus,
    title={Company-Names-Corpus},
    author={wainshine},
    year={2024},
    note ={\url{https://github.com/wainshine/Company-Names-Corpus}}
}

@article{liu2022novel,
  title={A Novel Methodology for Credit Spread Prediction: Depth-Gated Recurrent Neural Network with Self-Attention Mechanism},
  author={Liu, Xiao and Zhou, Rongxi and Qi, Daifeng and Xiong, Yahui},
  journal={Mathematical Problems in Engineering},
  volume={2022},
  number={1},
  pages={2557865},
  year={2022},
  publisher={Wiley Online Library}
}

@article{huang2025explainable,
  title={Explainable sentiment analysis with DeepSeek-R1: Performance, efficiency, and few-shot learning},
  author={Huang, Donghao and Wang, Zhaoxia},
  journal={IEEE Intelligent Systems},
  year={2025},
  publisher={IEEE}
}

@article{liu2023itransformer,
  title={itransformer: Inverted transformers are effective for time series forecasting},
  author={Liu, Yong and Hu, Tengge and Zhang, Haoran and Wu, Haixu and Wang, Shiyu and Ma, Lintao and Long, Mingsheng},
  journal={arXiv preprint arXiv:2310.06625},
  year={2023}
}

@inproceedings{zhou2021informer,
  title={Informer: Beyond efficient transformer for long sequence time-series forecasting},
  author={Zhou, Haoyi and Zhang, Shanghang and Peng, Jieqi and Zhang, Shuai and Li, Jianxin and Xiong, Hui and Zhang, Wancai},
  booktitle={Proceedings of the AAAI conference on artificial intelligence},
  volume={35},
  number={12},
  pages={11106--11115},
  year={2021}
}

@article{ansari2025chronos2,
  title        = {Chronos-2: From Univariate to Universal Forecasting},
  author       = {Abdul Fatir Ansari and Oleksandr Shchur and Jaris Küken and Andreas Auer and Boran Han and Pedro Mercado and Syama Sundar Rangapuram and Huibin Shen and Lorenzo Stella and Xiyuan Zhang and Mononito Goswami and Shubham Kapoor and Danielle C. Maddix and Pablo Guerron and Tony Hu and Junming Yin and Nick Erickson and Prateek Mutalik Desai and Hao Wang and Huzefa Rangwala and George Karypis and Yuyang Wang and Michael Bohlke-Schneider},
  journal      = {arXiv preprint arXiv:2510.15821},
  year         = {2025},
  url          = {https://arxiv.org/abs/2510.15821}
}

@inproceedings{Yuqietal-2023-PatchTST,
  title     = {A Time Series is Worth 64 Words: Long-term Forecasting with Transformers},
  author    = {Nie, Yuqi and
               H. Nguyen, Nam and
               Sinthong, Phanwadee and 
               Kalagnanam, Jayant},
  booktitle = {International Conference on Learning Representations},
  year      = {2023}
}

\appendix

\section{The Two-stage Labeling of \(\mathcal{D}_{1}\)}
\label{appd:label}

\begin{figure}[htp]
    \centering
    \includegraphics[width=0.95\linewidth]{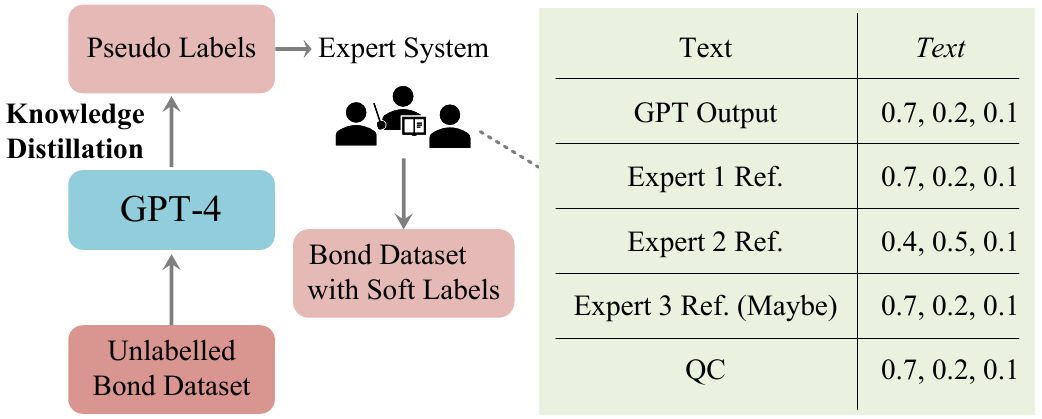}
    \caption{\textbf{The Two-stage Labeling Procedure of \(\mathcal{D}_{1}\)}.}
    \label{fig:label}
\end{figure}

The two-stage labeling for \(\mathcal{D}_{1}\) aims to accelerate annotation with GPT, with the procedure shown in Figure~\ref {fig:label}. In the first stage, GPT-4~\cite{achiam2023gpt} generates pseudo labels of coarse sentiment polarity probabilities (prompt details in Appendix~\ref{sec:prompts_absa}). In the second stage, our expert system rectifies them, and the rectification protocol is as follows: two experts first conduct the first round individually. If the absolute probability differences of polarities do not exceed $0.1$, the results are sent to a quality control (QC) expert for verification; otherwise, a third expert reviews their results before submission to the QC expert, ultimately yielding soft labels of sentiment polarity probabilities. 

Our experts consist of 7 Ph.D. candidates majoring in financial engineering. Among them, $1$ expert handles QC, who is the final decision maker who integrates experts' opinions. \(\mathcal{D}_{1}\) includes $6,881$ samples, for every $1,000$ samples, the other $6$ experts are randomly divided into groups of 3 to annotate. Below are some statistics (selecting the class with the highest probability): For the inter-rater agreement among experts, we use Fleiss' Kappa, computed using only Expert 1, Expert 2, and QC, since cases involving Expert 3 are very rare. The Kappa values for each slot are $0.8458$, $0.819$, $0.8420$, $0.8691$, $0.8598$, and $0.8413$. For the initial agreement between GPT and the final labels, we use Cohen's Kappa, which is $0.8644$.

In industry practice, one concern with GPT-assisted annotation is the potential introduction of anchoring: experts may be biased toward GPT-proposed probabilities, especially when these values appear plausible. To mitigate this issue, we conducted structured training for the annotation team before the two-stage labeling process:
\begin{itemize}
    \item The training mainly introduced the sentiment scale over the interval $[-1, 1]$, divided into slots of width $0.2$, where each slot corresponds to several typical expressions. For example, phrases such as ``marginal improvement in fundamentals'' or ``a recovery in market investment sentiment'' correspond to the interval $(0.2, 0.4]$. 
    \item Prior to formal annotation, each annotator was assigned a trial set of 100 samples without access to the original GPT outputs. Admission to the annotation team required passing this qualification phase, defined as achieving absolute probability differences of polarity within $0.1$ relative to the ground truth for each item.
\end{itemize}
After training, during the formal annotation of the 6{,}881 samples in $D_{1}$, there was no case in which the QC expert simultaneously rejected the judgments of both experts.

\section{Named Entity Recognition}
\label{appd:ner}

\begin{figure}[htp]
    \centering
    \includegraphics[width=1.0\linewidth]{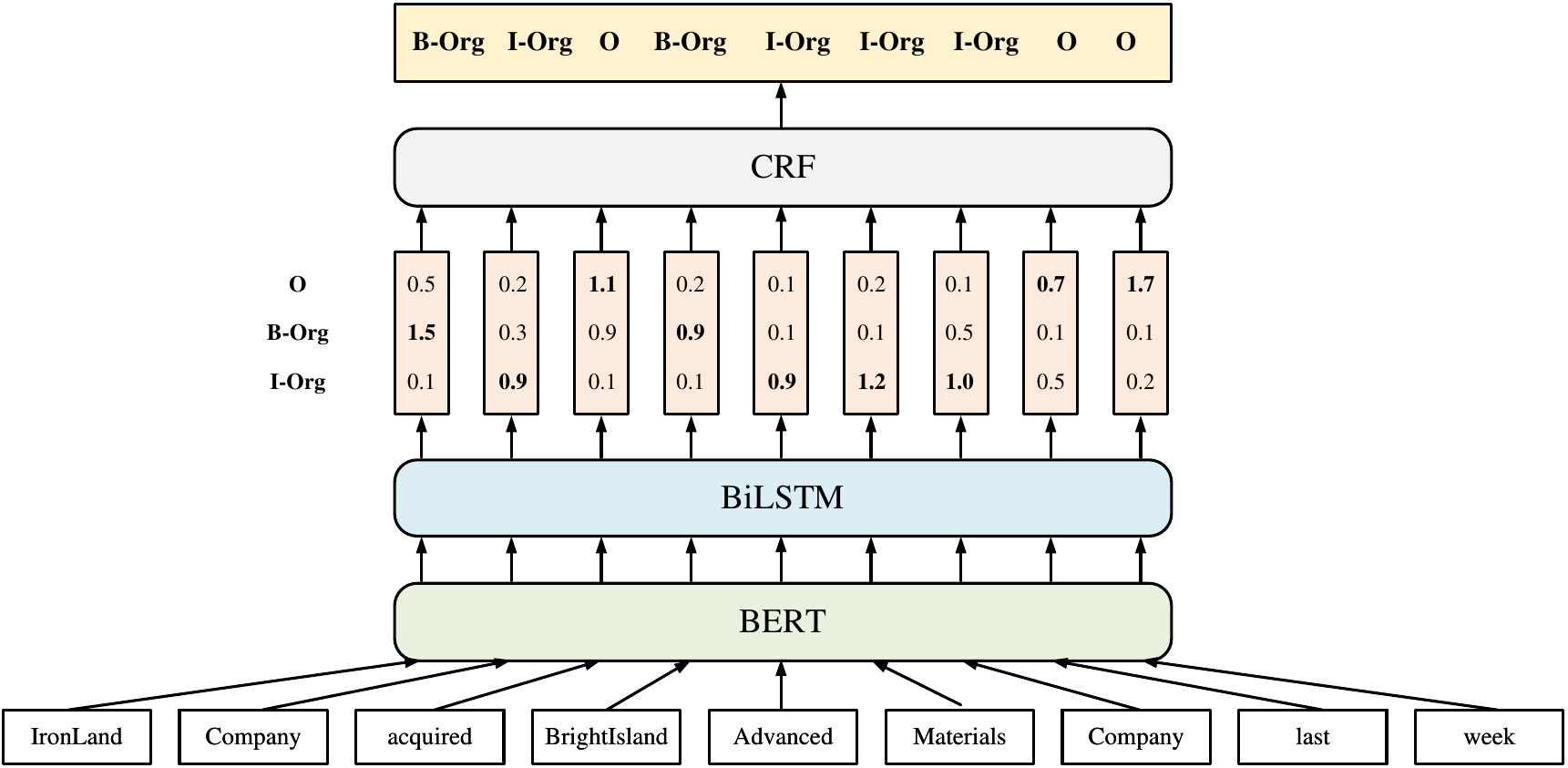}
    \caption{A General Introduction to Implementing NER.}
    \label{fig:ner}
\end{figure}

Named entity recognition (NER) is formulated as a sequence labeling task. The objective is to classify each entity in a natural language sentence \cite{ehrmann2023named}. We use the \href{https://github.com/wainshine/Company-Names-Corpus}{\textit{Company-Names-Corpus}} for supervised training and inference on $\mathcal{D}_{2}$. In this work, we categorize each entity into one of three classes: \{B-org, I-org, O\}, as illustrated in Figure~\ref{fig:ner}. The purpose is to assign texts containing identifiable firm entities to $\mathcal{D}_{2, \alpha}$, while texts without any firm entities are assigned to $\mathcal{D}_{2, \beta}$. The presence of firm entities suggests that the corresponding text is more suitable for micro-level ABSA, whereas texts without entities are more appropriate for meso-level SLSA. The process in Figure~\ref{fig:ner} can be described as follows: the text input is converted into token IDs via a tokenizer and then fed into BERT for feature extraction. Considering that concatenating the forward and backward layers of BiLSTM can yield richer contextual representations, the output vectors are further processed through BiLSTM to extract contextual features required for entity recognition. Finally, they are fed into the CRF layer for decoding to compute the optimal annotation sequence. We report the performance of this component in Table~\ref{tab:ner}.

\begin{table}[h]
    \small
    \centering
    \setlength\tabcolsep{4pt} 
    \captionsetup{width=\linewidth}
    \caption{NER Performance Results.}
    \label{tab:ner}
    \begin{tabular}{cc} 
    \toprule
    Metric & Performance (\%) \\ 
    \midrule
    Precision & 94.55 \\
    Recall & 95.31 \\
    F1-Score & 94.92 \\
    \bottomrule
    \end{tabular}
\end{table}

Since the same entity in a text may have different expressions, in some cases, users require entities with the same conceptual reference but different expressions to have the most accurate descriptions, which are generally official and formal. Therefore, we introduce an external entity knowledge base (distinct from the Knowledge Base $\mathcal{B}$, we directly use the \href{https://github.com/wainshine/Company-Names-Corpus}{\textit{Company-Names-Corpus}}), and users can choose to refer to this base for closed-domain entity linking: if there is no official name of the corresponding entity in the text, the output of the model is empty. For linking entities in the closed domain, we use edit distance to calculate similarity:
\begin{equation}
\small
\begin{aligned}
\text{sim}_{\mathcal{W}} &= \text{sim}_\mathcal{J} + l_p \lambda(1 - \text{sim}_\mathcal{J}) \\
\text{s.t.} \quad \text{sim}_\mathcal{J} &=
\begin{cases}
0, & \text{if } \tau = 0 \\
\frac{1}{3} \left( 
\frac{\tau}{|l_1|} 
+ \frac{\tau}{|l_2|} 
+ \frac{\tau - \eta}{\tau} 
\right), & \text{otherwise}
\end{cases}
\end{aligned}
\end{equation}
where $|l_1|$ and $|l_2|$ denote the lengths of strings $l_1$ and $l_2$, $\tau$ is the number of matching characters between the two strings, $\eta$ is half the number of transpositions, $l_p$ is the length of the common prefix of the two strings (with a maximum of $4$), and $\lambda$ is a scaling coefficient controlling the contribution of the common prefix.
Overall, the matching thresholds and tie-breaking rules are as follows: 
\begin{itemize}
    \item We first perform closed-domain entity linking, where the edit distance threshold is set to 0.8. Since company names in the \href{https://github.com/wainshine/Company-Names-Corpus}{\textit{Company-Names-Corpus}} are unique, tie-breaking does not arise at this stage.
    \item The number of bond and firm entities changes rapidly, if closed-domain entity linking fails, we proceed to open-domain entity disambiguation, where the edit distance threshold is also set to 0.8. In cases of tie-breaking, a set of mutually referring similar entities is extracted. In the Chinese context, these typically correspond to the full name and its abbreviated form. In such cases, we select the shorter abbreviated name as the final entity.
\end{itemize}

\section{Experiments Setting}
\label{appd:setting}

\subsection{Data Standardization} 

The order of magnitude of each feature column in $\mathcal{D}_{3}$ is inconsistent, thus we apply \textit{z-score} standardization to each feature column to accelerate the convergence for BDRF modeling.

\subsection{Hyperparameters} This work is implemented by \textit{PyTorch} on an NVIDIA RTX A5000 GPU.

For ABSA BERT followed by an MLP (consisting of three layers, with hidden dimensions of $768*3$, $768$ and $64$, respectively), we used MSE as the criterion, with \textit{Adam} optimizer ($\text{lr}=1\text{e-}4$, weight decay for BERT = $1\text{e-}5$ and weight decay for MLP = $1\text{e-}7$, respectively). Train for $50$ epochs (about 2 hours). During inference, we take the polarity with the highest probability among the three polarities as the output.  
\begin{equation}
\small
    MSE(\mathbf{p}_{s},\hat{\mathbf{p}}_{s})= \frac{1}{U}\sum_{u=1}^{U}(\mathbf{p}_{s}-\hat{\mathbf{p}}_{s})^{2} \quad ,
\end{equation}
where $U$ is the number of texts in $\mathcal{D}_{1}$, $\mathbf{p}_{s}$ is a three-dimensional vector indicating three polarities.

For the aggregation AttnMLP, both the initial linear layer and the output linear layer have dimension $1$, and the latent dimension is set to $16$. 
    
For the forecasting model, we use iTransformer. The \textit{seq\_len} is aligned with the window length $T$ and set to $21$. The \textit{pred\_len} equals the forecasting target $q$; for $t+1$ forecasting, \textit{pred\_len} is set to $1$. The model dimension $d_{\text{model}}$ is set to $64$, and the number of encoder layers $e_{\text{layers}}$ is set to 3. We used RMSE as the criterion, with \textit{RMSprop} optimizer ($\text{lr}=1\text{e-}4$, weight decay = $1\text{e-}7$, momentum =  $0.9$). Train for $50$ epochs (about 11 hours). 
    \begin{equation}
    \small
        RMSE(\y_{i},\hat{\y}_{i})= \sqrt{\frac{1}{N_{train}}\sum_{i=1}^{N_{train}}(\y_{i}-\hat{\y}_{i})^{2}} \quad ,
    \end{equation}
    where $N_{train}$ is the total number of bonds in $\mathcal{D}_{3}$ trainset, $\y_{i}$ is the \textit{credit spread} of bond $i$.

\section{Pseudocode of Mean-Max Pooling}
\label{algo:mean-max-pool}

\begin{algorithm}[!h]
    \small
    \SetAlgoLined
    \caption{Mean-Max Pooling}
    \label{pseudocode:mean-max-pool}
    \KwIn{labeled dataset $\mathcal{D}_{1}$, unlabeled dataset $\mathcal{D}_{2,\alpha}$, encoder $f_{1}(\cdot)$, classifier $c_{\alpha}(\cdot)$, bond entity $i$}
    \KwOut{bond-specific pooled sentiment $s^{\mathrm{mmp}}_{\alpha,i}$}
    
    $\mathcal{D}^{(\alpha)} \leftarrow \mathcal{D}_{1} \cup \mathcal{D}_{2,\alpha}$ \;
    
    \ForEach{text $j \in \mathcal{D}^{(\alpha)}$}{
        $(\mathbf{e}^{(j)}_{\mathrm{cls}}, \{\mathbf{e}^{(j)}_{r}\}_{r=1}^{L_j}) \leftarrow f_{1}(j)$ \;
        
        \If{$i \in j$}{
            $\mathcal{P}_{i,j} \leftarrow$ token positions of entity $i$ in text $j$ \;
            
            $\mathcal{E}_{i,j} \leftarrow \{\mathbf{e}^{(j)}_{r} \mid r \in \mathcal{P}_{i,j}\}$ \;
            
            $\mathbf{e}^{\mathrm{mean}}_{i,j} \leftarrow \mathrm{Mean}(\mathcal{E}_{i,j})$ \;
            $\mathbf{e}^{\mathrm{max}}_{i,j} \leftarrow \mathrm{MaxPool}(\mathcal{E}_{i,j})$ \;
            
            $\mathbf{z}^{\mathrm{mmp}}_{i,j} \leftarrow [\mathbf{e}^{(j)}_{\mathrm{cls}} ; \mathbf{e}^{\mathrm{mean}}_{i,j} ; \mathbf{e}^{\mathrm{max}}_{i,j}]$ \;
            
            $s^{\mathrm{mmp}}_{\alpha,i,j} \leftarrow c_{\alpha}(\mathbf{z}^{\mathrm{mmp}}_{i,j})$ \;
        }
    }
    
    $s^{\mathrm{mmp}}_{\alpha,i} \leftarrow \max_{j \in \mathcal{D}^{(\alpha)},\, i \in j} s^{\mathrm{mmp}}_{\alpha,i,j}$ \;
    
    \Return{$s^{\mathrm{mmp}}_{\alpha,i}$}
\end{algorithm}

\section{Metrics Calculation}
\label{appd:metrics}
\begin{equation}
\small
    Precision = \frac{TP_{s}}{TP_{s}+FP_{s}}\quad ,
\end{equation}
where $TP_{s}$ and $FP_{s}$ are the numbers of correct and incorrect polarities in the $\mathcal{D}_{1}$ testset, respectively.

\begin{equation}
\small
    MAE(\y_{i},\hat{\y}_{i}) = \frac{1}{N_{test}}\sum_{i=1}^{N_{test}}|\y_{i}-\hat{\y}_{i}|\quad ,
\end{equation}
\begin{equation}
\small
    MAPE(\y_{i},\hat{\y}_{i}) = \frac{1}{N_{test}}\sum_{i=1}^{N_{test}}|\frac{\y_{i}-\hat{\y}_{i}}{\y_{i}}|\quad ,
\end{equation}
where $N_{test}$ is the total number of bonds in $\mathcal{D}_{3}$ testset, $\y_{i}$ is the credit spread of bond $i$.

\section{Pseudocode of RAG Mapping}
\label{algo:rag}

\begin{algorithm}[!h]
    \small
    \SetAlgoLined
    \caption{RAG-based Mapping}
    \label{pseudocode:rag}
    \KwIn{unlabeled dataset $\mathcal{D}_{2,\beta}$, LLM agent $f_{2}(\cdot)$, embedding model $f_{3}(\cdot)$, topic set $\mathcal{B}'$, weighting function $\psi_{\beta}(\cdot)$}
    \KwOut{RAG-based meso-level sentiment $s^{\mathrm{rag}}_{\beta}$}
    
    $s^{\mathrm{rag}}_{\beta} \leftarrow 0$ \;
    
    \ForEach{text $j \in \mathcal{D}_{2,\beta}$}{
        $y_{j} \leftarrow f_{2}(j)$ \tcp*{\textcolor{gray}{text sentiment}}
        $\mathbf{v}_{j} \leftarrow f_{3}(j)$ \tcp*{\textcolor{gray}{text embedding}}
        
        \ForEach{topic $n \in \mathcal{B}'$}{
            $\mathbf{v}_{n} \leftarrow f_{3}(n)$ \;
            $c_{j,n} \leftarrow \cos(\mathbf{v}_{j}, \mathbf{v}_{n})$ \;
        }
        
        $N^{\ast} \leftarrow \text{TopK}(\{c_{j,n}\}_{n \in \mathcal{B}'}, 5)$ \tcp*{\textcolor{gray}{top-5 nearest topics}}
        
        $a_{j} \leftarrow \psi_{\beta}(N^{\ast})$ \tcp*{\textcolor{gray}{aggregation weight}}
        
        $s^{\mathrm{rag}}_{\beta} \leftarrow s^{\mathrm{rag}}_{\beta} + a_{j} \cdot y_{j}$ \;
    }
    
    \Return{$s^{\mathrm{rag}}_{\beta}$}
\end{algorithm}

\section{Results on ABSA}
\label{exp:micro}

\begin{table}[h]
    \small
    \centering
    \setlength\tabcolsep{4pt} 
    \captionsetup{width=\linewidth}
    \caption{Pooling Comparison Results.}
    \label{tab:pool}
    \begin{tabular}{cc} 
    \toprule
    Method & Precision (\%) \\ 
    \midrule
    Mean Pooling  & 86.19  \\    
    Max Pooling  & 87.59  \\    
    \textbf{Mean-Max Pooling}  & \textbf{88.27}  \\  
    \bottomrule
    \end{tabular}
\end{table}

Standard ABSA is typically formulated as a sequence labeling problem, which predicts sentiment categories such as \{POS, NEG, NEU\}~\cite{gong2020unified}. In contrast, our objective is to end-to-end output entity-level sentiment in numerical form. Moreover, \(\mathcal{D}_{1}\) provides soft labels for specific entities, rather than token-level BIO-style annotations. In our trials, this data format was not suitable for training sequence labeling models. Therefore, standard ABSA approaches cannot be directly used as baselines in our setting. In addition to the mean-max pooling method described in Appendix~\ref{algo:mean-max-pool} for extracting global features, we attempted two alternative baselines by directly concatenating the results of mean pooling and max pooling with the $[CLS]$ embedding. The results are in Table~\ref{tab:pool}, which shows that the mean-max pooling achieves the best performance among the three methods.

\section{Pseudocode of BDRF Modeling}
\label{algo:bdp}

$\mathbf{x}_{i,k}$ is the feature vector including sentiment $s$ for bond $i$ at time step $k$, $\mathbf{y}_{i}$ is the credit spread sequence of bond $i$, and $I$ is the total number of bonds in $\mathcal{D}_{3}$. 

\begin{algorithm}[h]
    \small
    \SetAlgoLined
    \caption{BDRF Modeling for Credit Spread Forecasting}
    \label{pseudocode:bdp}
    \KwIn{rolling window length $T$, bond feature dataset $\mathcal{D}_{3}$, model $f_{4}(\cdot)$, forecasting horizon $q$}
    \KwOut{\textit{credit spread} sequences $\{\hat{\mathbf{y}}_{i}\}_{i=1}^{I}$}
    
    {\color{gray}
    \tcp{$\mathcal{D}_{3} = \{(\{\mathbf{x}_{i,k}\}_{k=1}^{K}, \mathbf{y}_{i})\}_{i=1}^{I}$}
    }
    
    \For{$i = 1$ \KwTo $I$}{
        
        \For{$k = 1$ \KwTo $K - T - q$}{
            
            {\color{gray}
            \tcp{$k$ indexes time steps; window size is $T$}
            \tcp{$q$ denotes the forecasting horizon (e.g., $q=1,2$)}
            }
            
            $\mathbf{X}^{(T)}_{i,k} \leftarrow \{\mathbf{x}_{i,k}, \dots, \mathbf{x}_{i,k+T-1}\}$ \;
            
            $\hat{y}_{i,k+q} \leftarrow f_{4}(\mathbf{X}^{(T)}_{i,k})$ \;
        }
        
        $\hat{\mathbf{y}}_{i} \leftarrow \{\hat{y}_{i,k+q}\}_{k=1}^{K - T - q}$ \;
    }
    
    \Return{$\{\hat{\mathbf{y}}_{i}\}_{i=1}^{I}$}
\end{algorithm}

\section{Results on SLSA}
\label{exp:meso}

\footnotetext[1\label{fn:llama}]{\url{https://huggingface.co/hfl/llama-3-chinese-8b}}

\begin{table}[htp]
    \tiny
    \centering
    \setlength\tabcolsep{4pt} 
    \captionsetup{width=\linewidth}
    \caption{Foundation Model Comparison Results.}
    \label{tab:llm}
    \begin{tabular}{cc} 
    \toprule
    Foundation Model & Precision (\%) \\ 
    \midrule
    \textit{DeepSeek-R1-Distill-Qwen-1.5B}~\cite{deepseekai2025deepseekr1incentivizingreasoningcapability}  & 51.8  \\
    \textit{Llama-3-Chinese-8B}\textsuperscript{\hyperref[fn:llama]{1}}  & 55.4  \\ 
    \textit{Baichuan-7B}~\cite{yang2023baichuan} & 61.5  \\
    \textit{DeepSeek-R1-Distill-Qwen-7B}~\cite{deepseekai2025deepseekr1incentivizingreasoningcapability}  & 65.7  \\
    \textit{XuanYuan-6B}~\cite{zhang2023xuanyuan20largechinese} & 68.9 \\
    \textit{GPT-4}~\cite{achiam2023gpt} & 72.0 \\
    \textbf{\textit{Qwen2.5-3B-Instruct}}~\cite{qwen2.5}  & \textbf{75.0}  \\ 
    \textit{Qwen2.5-72B}~\cite{qwen2.5} & 75.8 \\
    \textit{DeepSeek-R1-32B}~\cite{deepseekai2025deepseekr1incentivizingreasoningcapability} & 76.2 \\
    \bottomrule
    \end{tabular}
\end{table}

In this experiment, we evaluated $7$ foundation models under the same prompt settings. Since $\mathcal{D}_{2,\beta}$ has 723,260 entries, we randomly sampled $5$ subsets of 100 entries each. The models independently generated sentiment outputs for these subsets, and then our financial experts assessed their precision, the average precision across the $5$ subsets was taken as the model's performance. Prior studies have shown that for discriminative tasks such as sentiment classification, the conventional belief that larger models yield better performance does not always hold~\cite{huang2025explainable}. Considering inference cost and deployment feasibility (our meso-level module processes 723,260 texts), we primarily adopt small-scale model Qwen2.5-3B-Instruct (75.0\%), which provides the best trade-off between performance and computational efficiency.

\section{Feature Set Taxonomy} 
\label{appd:firms_features}
The dependent variable of the BDRF model $f_4$ is the \textit{credit spread}, while the independent variables include the numeric features listed in Table~\ref{tab:feature_set} and our sentiment. All numeric independent variables are \textit{z-score} normalized before being input into $f_4$.

\renewcommand{\thefootnote}{\arabic{footnote}}
 
\begin{table}[htp]
    \tiny
    \centering
    \setlength\tabcolsep{3pt} 
    \captionsetup{width=\linewidth}
    \caption{Feature Set Taxonomy.}
    \label{tab:feature_set}
    \begin{tabular}{c|p{0.6\linewidth}}  
    \toprule
    Dimension & \multicolumn{1}{c}{Feature Name} \\  
    \midrule
    \multirow{9}{*}{\centering \makecell{Macroeconomic and \\Financial Indicators}} 
    & 1. USDCNYC  \\  
    & 2. Shibor (Shanghai Interbank Offered Rate) in March \\  
    & 3. Manufacturing PMI (Purchasing Managers' Index) \\  
    & 4. Macroeconomic Prosperity Index: Leading Index \\  
    & 5. PPI (Producer Price Index): Year-over-Year for the Current Month \\  
    & 6. GDP (Gross Domestic Product): Year-over-Year for the Current Quarter \\ 
    & 7. CPI (Consumer Price Index): Year-over-Year for the Current Month \\ 
    & 8. Aggregate Financing to the Real Economy (AFRE): Year-over-Year at Period-End \\
    & 9. Yield on Government Bonds (for the Corresponding Period) \\
    \midrule
    \multirow{1}{*}{\centering \makecell{Industrial Indicator}} 
    & 10. \href{https://www.swsresearch.com/institute_sw/home}{\textit{SWS}} Primary Industry Index \\
    \midrule
    \multirow{1}{*}{\centering \makecell{Trading Indicator}} 
    & 11. Trading Volume \\  
    \midrule
    \multirow{31}{*}{\centering \makecell{Firm Financial \\ and Operational \\ Indicators}} 
    & 12. Operating Revenue \\  
    & 13. Operating Costs \\  
    & 14. Total Profit \\  
    & 15. Current Assets \\  
    & 16. Non-Current Assets \\  
    & 17. Total Assets \\  
    & 18. Current Liabilities \\  
    & 19. Non-Current Liabilities \\  
    & 20. Total Liabilities \\  
    & 21. Total Shareholders’ Equity \\  
    & 22. Cash Flow from Operations \\  
    & 23. Cash Flow from Investment \\  
    & 24. Cash Flow from Finance \\  
    & 25. Total Cash Flow \\  
    & 26. Current Ratio \\  
    & 27. Quick Ratio \\  
    & 28. Super Quick Ratio \\  
    & 29. Debt-to-Asset Ratio (\%) \\  
    & 30. Equity Ratio (\%) \\  
    & 31. Tangible Net Worth Debt Ratio (\%) \\  
    & 32. Gross Profit Margin (\%) \\  
    & 33. Net Profit Margin (\%) \\  
    & 34. Return on Assets (\%) \\  
    & 35. Operating Profit Margin (\%) \\  
    & 36. Average Return on Equity (\%) \\  
    & 37. Operating Cycle (Days) \\  
    & 38. Inventory Turnover Ratio \\  
    & 39. Accounts Receivable Turnover Ratio \\  
    & 40. Current Asset Turnover Ratio \\  
    & 41. Shareholders’ Equity Turnover Ratio \\  
    & 42. Total Asset Turnover Ratio \\  
    \midrule
    \multirow{3}{*}{\centering \makecell{Firm Comprehensive \\ Credit Indicators}} 
    & 43. Remaining Credit Utilization Ratio \\  
    & 44. Month-over-Month Change in Credit \\  
    & 45. Secured Credit Ratio \\  
    \bottomrule
    \end{tabular}
\end{table}

\section{The Correlation Matrix of Features}
\label{appd:corr mat}

\begin{figure}[!h]
    \centering
    \includegraphics[width=1.0\linewidth]{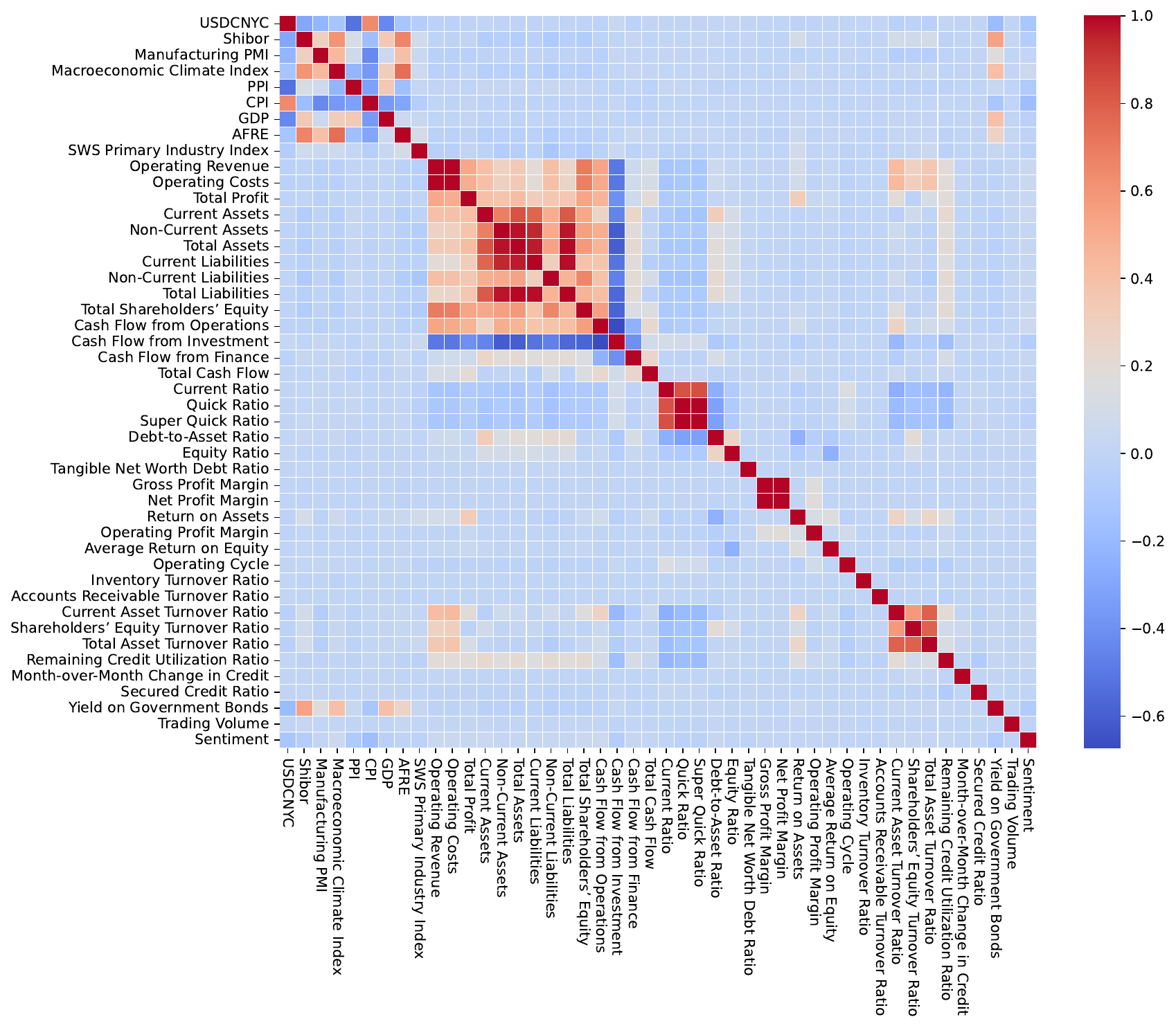}
    \caption{The Correlation Matrix of Features.}
    \label{fig:corr_mat}
\end{figure}

As in Figure~\ref{fig:corr_mat}, strong collinearity exists within the \textit{Firm Financial and Operational Indicators} listed in Appendix~\ref{appd:firms_features} among the remaining independent variables. Within the \textit{Macroeconomic and Financial Indicators}, collinearity is present among \textit{Shibor}, \textit{Manufacturing PMI}, \textit{Macroeconomic Climate Leading Index}, \textit{GDP}, and \textit{AFRE}.

\section{Feature Attribution}
\label{appd:feature_importance}

The results are shown in Table~\ref{tab:feature_importance}, where the feature importance is calculated by randomly permuting the positions of independent variable features \cite{altmann2010permutation}. The composite sentiment we extracted ranks $17$th among the $46$ features.

\begin{table}[!h]
    \small
    \centering
    \setlength\tabcolsep{3pt} 
    \captionsetup{width=\linewidth}
    \caption{Feature Attribution.}
    \label{tab:feature_importance}
    \begin{tabular}{l|c} 
    \toprule
    Feature Name & Feature Importance \\ 
    \midrule
1: PPI & 0.05589 \\
2: Macroeconomic Climate Index & 0.05012 \\
3: AFRE & 0.04865 \\
4: Yield on Government Bonds & 0.04769 \\
5: GDP & 0.04218 \\
6: Shibor & 0.03431 \\ 
7: CPI & 0.02487 \\ 
8: Cash Flow from Finance & 0.02349 \\
\textbf{9: \textit{Sentiment}} & \textbf{0.02245} \\
10: Manufacturing PMI & 0.02241 \\
11: Current Ratio & 0.01738 \\
12: Total Asset Turnover Ratio & 0.01671 \\
13: Total Cash Flow & 0.01322 \\
14: Return on Assets & 0.01049 \\ 
15: Cash Flow from Investment & 0.00891 \\
16: Remaining Credit Utilization Ratio & 0.00842 \\
17: USDCNYC & 0.00836 \\ 
18: Current Asset Turnover Ratio & 0.00758 \\
19: SWS Primary Industry Index & 0.00753 \\
20: Super Quick Ratio & 0.00714 \\
21: Shareholders’ Equity Turnover Ratio & 0.00655 \\
22: Non-Current Liabilities & 0.00637 \\
23: Cash Flow from Operations & 0.00605 \\
24: Operating Cycle & 0.00579 \\
25: Total Shareholders’ Equity & 0.00535 \\ 
26: Current Assets & 0.00512 \\ 
27: Operating Costs & 0.00494 \\
28: Equity Ratio & 0.00473 \\ 
29: Total Profit & 0.00426 \\
30: Quick Ratio & 0.00412 \\
31: Trading Volume & 0.00379 \\
32: Operating Revenue & 0.00381 \\
33: Operating Profit Margin & 0.00366 \\
34: Current Liabilities & 0.00269 \\ 
35: Debt-to-Asset Ratio & 0.00197 \\ 
36: Tangible Net Worth Debt Ratio & 0.00178 \\
37: Non-Current Assets & 0.00123 \\
38: Average Return on Equity & 0.00106 \\
39: Gross Profit Margin & 0.00088 \\
40: Total Assets & 0.00079 \\
41: Net Profit Margin & 0.00062 \\
42: Total Liabilities & 0.00044 \\
43: Accounts Receivable Turnover Ratio & 0.00036 \\
44: Month-over-Month Change in Credit & 0.00017 \\
45: Secured Credit Ratio & 0.00014 \\
46: Inventory Turnover Ratio & 1.73e-08 \\
    \bottomrule
    \end{tabular}
\end{table}

\section{Visualization Explanation}

\subsection{Explanation on Figure~\ref{fig:year_line}}
\label{appd:exp on fig3}
In Figure~\ref{fig:year_line}, we highlighted several time slots and found that they correspond to influential events in Chinese society. For example, \ding{172} refers to the mid-2015 \textit{Chinese Stock Market Turbulence}, during which a large number of retail investors, affected by short-selling and leverage mechanisms, began to sell off stocks. As a result, the bond market was severely impacted, and sentiment across all industries remained depressed for several months. \ding{178} corresponds to the end of 2022, when China abandoned the ``\textit{Zero-COVID Policy}.'' The economy, previously constrained by the pandemic, began to recover, leading to a shift in sentiment from bearish to bullish across most industries, whereas industries such as \textit{Television and Broadcasting}, \textit{Gaming}, \textit{Advertising and Marketing}, \textit{Digital Media}, \textit{Social Media}, and \textit{Publishing} remained depressed, as people had fewer opportunities to socialize through these media while staying at home.  

\subsection{Explanation on Figure~\ref{fig:decomposition} a)}
To analyze whether the industry sentiment time series exhibits seasonal trends, we selected the industry \textit{Automobile} and performed decomposition using \textit{Seasonal Decomposition of Time Series by Loess (STL)}. The visualization is in Figure~\ref{fig:decomposition} a). The third subplot, ``Season,'' indicates that industry sentiment does not exhibit significant seasonality. The fourth subplot, ``Resid,'' shows that the noise distribution remains stable, with few outliers.

\section{Dataset Structure}
\label{appd:stats}

The data structures of $\mathcal{D}_{1}$, $\mathcal{D}_{2,\alpha}$, $\mathcal{D}_{2,\beta}$, and $\mathcal{D}_{3}$ are shown in Table~\ref{tab:data_structure}, with italic font marking fixed keys. In $\mathcal{D}_{3}$, italicized features refer to those in Table ~\ref{tab:feature_set}.

\begin{table}[!h]
    \small
    \centering
    \setlength\tabcolsep{4pt} 
    \captionsetup{width=\linewidth}
    \caption{Metadata Structure.}
    \label{tab:data_structure}
    \begin{tabular}{c|p{0.6\linewidth}}  
    \toprule
    Dataset & \multicolumn{1}{c}{Key-value Pair Description} \\  
    \midrule
    \multirow{13}{*}{\centering \makecell{$\mathcal{D}_{1}$}} 
    & [ \\
    & \quad \{ \\
    & \quad \quad \textit{Date}: A timestamp.  \\  
    & \quad \quad \textit{MediumSource}: The name of the medium.  \\  
    & \quad \quad \textit{Title}: This is the title.  \\  
    & \quad \quad \textit{Content}: This is the content. \\  
    & \quad \quad \textit{Firm}: The name of the firm.  \\  
    & \quad \quad \textit{Tokens}: An integer.  \\  
    & \quad \quad \textit{NegProb}: A float in $[0,1]$.  \\  
    & \quad \quad \textit{NeuProb}: A float in $[0,1]$. \\  
    & \quad \quad \textit{PosProb}: A float in $[0,1]$. \\  
    & \quad \} \\
    & ] \\
    \midrule
    \multirow{8}{*}{\centering \makecell{$\mathcal{D}_{2,\alpha}$}} 
    & \{ \\
    &\quad Firm Name:  \\
    &\quad \{  \\
    &\quad \quad Date 1: [Text 1, Text 2, ...]  \\ 
    &\quad \quad Date 2: [Text 1, Text 2, ...]  \\
    &\quad \quad ...  \\
    &\quad \}  \\
    & \} \\
    \midrule
    \multirow{11}{*}{\centering \makecell{$\mathcal{D}_{2,\beta}$}} 
    & [ \\
    & \quad \{ \\
    & \quad \quad \textit{Date}: A timestamp.  \\  
    & \quad \quad \textit{MediumSource}: The name of the medium.  \\  
    & \quad \quad \textit{SerialNumber}: This is the number.  \\  
    & \quad \quad \textit{Section}: This is its section. \\  
    & \quad \quad \textit{Reporter}: The name of the reporters.  \\  
    & \quad \quad \textit{Title}: This is the title.  \\  
    & \quad \quad \textit{Body}: This is the body.  \\  
    & \quad \} \\
    & ] \\
    \midrule
    \multirow{8}{*}{\centering \makecell{$\mathcal{D}_{3}$}} 
    & [ \\
    &\quad \{ \\
    &\quad \quad \textit{Feature Name 1}: A float. \\
    &\quad \quad \textit{Feature Name 2}: A float. \\
    &\quad \quad ... \\
    &\quad \quad \textit{Credit Spread}: A float. \\
    &\quad \}  \\
    & ] \\
    \bottomrule
    \end{tabular}
\end{table}

\section{Experiments on Duration Function}
\label{appd:duration}

\begin{table}[!h]
    \centering
    \captionsetup{width=\linewidth}
    \caption{Comparison of different Duration Functions $h(\cdot)$ in the BDRF model.}
    \label{tab:exp_duration}
    \resizebox{\linewidth}{!}{
        \setlength\tabcolsep{2pt}
        \begin{tabular}{cccccc} 
        \toprule
        Duration Function & \makecell{MAE\\(e-5)} & \makecell{MAPE\\(e-3)} & $p$ 
        & \makecell{$\Delta$MAE\\(\%$\downarrow$)} & \makecell{$\Delta$MAPE\\(\%$\downarrow$)} \\
        \midrule
        n/a & 8.39 & 8.29 & n/a & n/a & n/a \\
        \makecell{\textit{Exponential Smoothing} \\ ($\alpha$=0.3)} & 8.44 & 8.13 & 0.054 & -0.59 & 1.93 \\
        \makecell{\textit{Smoothing Spline} \\ (factor=16)} & 8.21 & 8.03 & 0.039 & 2.14 & 3.13 \\ 
        \makecell{\textit{Daubechies 4} \\ (level=3)} & 7.94 & 8.47 & 0.052 & 5.36 & -2.17 \\ 
        \makecell{\textit{\textbf{Daubechies 4}} \\ (\textbf{level=6, $f^{*}$})} & \textbf{7.53} & \textbf{7.30} & \textbf{0.041} & \textbf{10.25} & \textbf{11.94} \\ 
        \bottomrule
        \end{tabular}
    }
\end{table}

To validate the rationality of our chosen duration function $h(\cdot)$, we conducted a comparative experiment on different functions, with the results shown in Table~\ref{tab:exp_duration}. The first row shows the results without sentiment, i.e., without applying any duration function. Compared to the other three baselines, our approach obtains more robust results under statistically significant \(p\)-values.  

\section{Case Study}
\label{appd:case}

\begin{figure}[!h]
    \centering
    \includegraphics[width=1.0\linewidth]{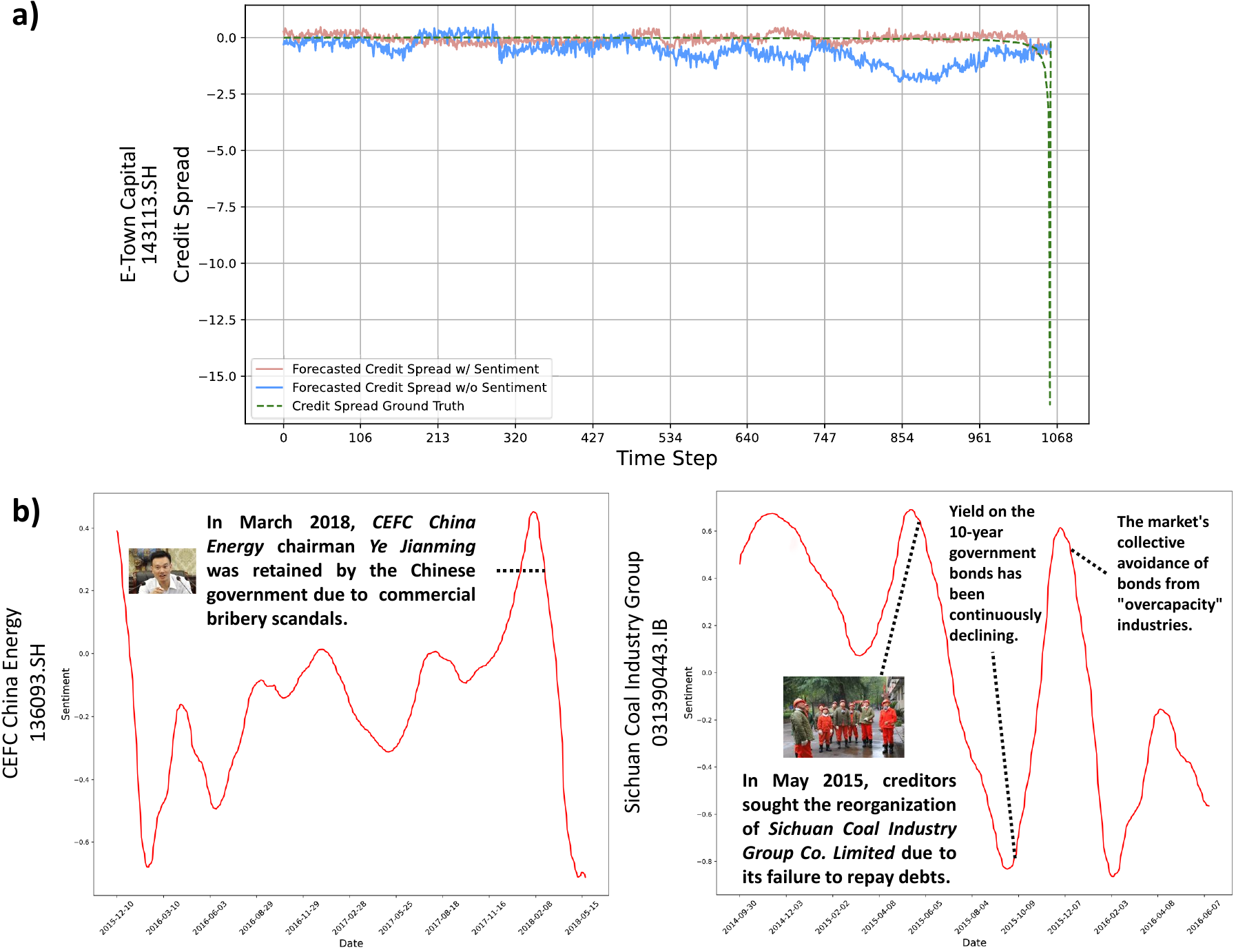}
    \caption{\textbf{a) Visualization of The Difference in Forecasting. b) Visualization of Composite Sentiment Dynamics of Defaulted Bonds Preceding defaults}. ``SH'' indicates the Shanghai Stock Exchange, and ``IB'' indicates the Investment Bank.}
    \label{fig:BDRF_viz}
\end{figure}

Having demonstrated the sentiment-risk event association at the industry level in Figure~\ref{fig:year_line}, we visualize it at the firm level with results in Figure~\ref{fig:BDRF_viz} a) and b), particularly pre-default composite sentiment changes: (1) \textit{credit spread} trajectory differences (with vs. without our sentiment) for one defaulted bond; (2) sentiment trajectories for two other defaulted bonds. For a), the sentiment-included trajectory is more stable than the sentiment-excluded one, which fluctuates sharply in most cases, and avoids contradicting the ground truth, unlike the sentiment-excluded one, which even produces incorrect forecasting of sustained market bullishness before the bond default. For b), both bonds showed significant pre-default sentiment shifts. For Bond \textit{136093.SH} issued by \textit{CEFC China Energy}, in March 2018, its chairman \textit{Ye} was detained over government official bribery scandals. With the board paralyzed and public outcry intensifying, the company suspended bond trading and defaulted on repayments. For Bond \textit{031390443.IB} issued by \textit{Sichuan Coal Industry Group}, plagued by long-term operational issues and energy sector structural changes, creditors filed for reorganization in May 2015 after a debt default, triggering prolonged bearish sentiment. Although the situation temporarily improved at the end of 2015 due to a decline in the yield on 10-year government bonds, the company ultimately defaulted amid continued financial strain and structural adjustment policies, accompanied by a renewed downturn in sentiment.

\section{Prompt in ABSA}
\label{sec:prompts_absa}
\begin{tcolorbox}
\scriptsize
{**Task Description:**  

I will provide a piece of text related to a specific bond, then you need to determine the probabilities of three sentiment polarities: pessimistic, neutral, and optimistic. Ensure that the sum of these probabilities is 1, and maximize the difference between them to reflect the dominant sentiment in the text.}


{**Sentiment Polarities Definition:**  

1. Pessimistic: The text indicates a negative market expectation of declining bond yields, and severe operational issues within the firm.  

2. Neutral: The text does not provide a clear stance on market prospects or firm performance.

3. Optimistic: The text suggests that the bond has growth potential, with stable or rising expected yields, and a positive outlook for future development.}  


{**Requirements:**  

1. Only return the probability values for the three sentiment categories without any explanations or reasoning process.  

2. Follow the output format below:  }


{**Example Format:**

**Input:**  

Entity: [Bond Name]

Text: [Provided text content]

**Output:**  

[Bond Name], Pessimistic: X, Neutral: Y, Optimistic: Z

where X, Y, and Z represent the probabilities of each sentiment category, ensuring their sum is 1 and that the differences between them are significant.  }


{**Example:**  

**Input:**  

Entity: [Blu-ray
]
Text: On 2nd July, according to an announcement from the Shanghai Stock Exchange, a bond issued by Sichuan Blu-ray Development Co., Ltd. experienced abnormal trading and was suspended for half an hour.

**Output:**  

[Blu-ray], Pessimistic: 0.6, Neutral: 0.3, Optimistic: 0.1}
\end{tcolorbox}

\section{Prompt in SLSA}
\label{sec:prompt}

\begin{tcolorbox}
\scriptsize
{**Task Description:**  

You are a professional bond market analysis expert. You will receive a series of texts related to the bond market. Your task is to determine the sentiment for each text based on its content. The sentiment has three discrete labels:

1. Pessimistic (-1): The text describes factors reflecting negative market sentiment in macro-financial contexts, particularly the bond market. These include concerns about economic fundamentals (such as slowdowns or rising unemployment), expectations of tighter monetary policy and higher interest rates, reduced market liquidity, and risk events like defaults or downgrades.

   
   
   
   
   
2. Neutral (0): The text describes neutral market sentiment, characterized by stable economic fundamentals, unchanged monetary policy with priced-in expectations, balanced market liquidity, and absence of major risk events. The content is mostly unrelated to macro financial markets, especially the bond market.  

   
   
   
   
   
3. Optimistic (1): The text reflects positive market sentiment, with mentions of stronger economic fundamentals, expectations of rate cuts or accommodative policy, ample liquidity, and easing risk events such as debt relief or rating upgrades. It is related to macro financial markets, particularly the bond market.

   
   
   
   
   


**Output Requirements:**  

- a) First, identify the subject of the news and determine whether it is related to the bond market, economic fundamentals, interest rate policies, market liquidity, etc.  

- b) Then, judge whether the news contains positive, negative, or neutral market factors. 

- c) Combine the keywords (e.g., ‘improvement’, ‘rise’ = 1; ‘deterioration’, ‘default’ = -1; ‘stable’, ‘in line with expectations’ = 0) to make the final judgment.  

- d) Finally, return a single sentiment score with values in {-1, 0, 1}.  

- e) Do not output any additional content or explanations.  


Below are some examples of input-output cases:

\par\vspace{4pt}  
**Example 1:**  

Input: The Minister of Industry and Information Technology, Jin Zhuanglong, stated at the 2023 China Pharmaceutical Industry Development Conference that China has over 10,000 large pharmaceutical enterprises, contributing 4\% of industrial added value and producing 40\% of global bulk raw materials. China ranks second globally in new drug research, with many innovative drugs, key vaccines, traditional Chinese medicine preparations, and high-end medical devices approved, strengthening supply security and supporting public health and COVID-19 response.

Thought Process: The government published positive data about the pharmaceutical industry, reflecting a positive market outlook.

Output: 1  

\par\vspace{4pt}  
**Example 2:**  

Input: Shanghai Pudong Police Station issued the first batch of unmanned driving equipment identification plates, designed in light blue and white, with a regional abbreviation, letters, and numbers, marked "unmanned equipment" at the top. Some citizens linked it to autonomous vehicles. On November 15, Pudong authorities and enterprises clarified that it applies to low-speed unmanned delivery equipment, not autonomous cars. 

Thought Process: The government issued the unmanned driving equipment identification plate, but the clarification means the management is strengthened, and not as anticipated by the public. The news is balanced and neutral.

Output: 0  

\par\vspace{4pt}  
**Example 3:**  

Input: The Shanghai Central Meteorological Observatory issued a blue wind warning at 08:20 on November 17, 2024, forecasting strong cold air with maximum gusts of 7-8 in inland areas and 8-9 along the river and coastal areas in the next 24 hours.  

Thought Process: Extreme weather events are expected, which disrupt production and daily life, leading to a negative market outlook.

Output: -1  
}
\end{tcolorbox}

\section{A Subset of The Knowledge Graph $\mathcal{G}$ And The Knowledge Base $\mathcal{B}$}
\label{appd:knowledge_graph}

Since the entire $\mathcal{G}$ is too large to be displayed in full, we provide an example subset, as shown in Table~\ref {tab:graph_exsample}, which illustrates the impact of $6$ natural disaster topics on 40 industries. The definition of the topic \textit{Natural Disasters} in $\mathcal{B}$ is provided in the box below. The complete $\mathcal{G}$ and $\mathcal{B}$ can be accessed from our code repository. Definitions in \(\mathcal{B}\) are retrieved by the expert system from Wikipedia, and together with \(\mathcal{G}\), they are not constructed using the same two-stage procedure as \(\mathcal{D}_{1}\). Instead, it is developed through close discussion among 3 senior Ph.D. candidates from our annotation team of 7 Ph.D. candidates. 

Since validating \(\mathcal{G}\) is challenging due to its reliance on the subjective consensus of experts, while its reliability directly bounds the reliability of \(s_{\beta}\), we conducted a sensitivity analysis by randomly flipping entries in \(\mathcal{G}\) from $0$ to $1$ or from $1$ to $0$ at various corruption levels, and then re-ran the BDRF forecasting experiment with \(q=1\) and \(T=21\):

\begin{table}[!h]
    \centering
    \captionsetup{width=\linewidth}
    \caption{Sensitivity analysis of \(\mathcal{G}\).}
    \label{tab:exp_kg_sensitivity}
    \resizebox{\linewidth}{!}{
        \tiny
        \setlength\tabcolsep{3pt}
        \begin{tabular}{ccccc}
        \toprule
        \makecell{Corruption\\Level} 
        & \makecell{MAE\\(e-5)} 
        & \makecell{MAPE\\(e-3)} 
        & \makecell{\(\Delta\)MAE\\(\%\(\uparrow\))} 
        & \makecell{\(\Delta\)MAPE\\(\%\(\uparrow\))} \\
        \midrule
        0\% & \textbf{7.53} & \textbf{7.30} & n/a & n/a \\
        25\% & 12.49 & 10.38 & 65.87 & 42.19 \\
        50\% & 20.91 & 17.34 & 177.69 & 137.53 \\
        75\% & 27.94 & 25.10 & 271.05 & 243.84 \\
        100\% & 30.38 & 24.09 & 303.45 & 230.00 \\
        \bottomrule
        \end{tabular}
    }
\end{table}

The results show that the forecasting performance degrades monotonically as the corruption level increases. Under 25\% random corruption, the MAE increases from 7.53 to 12.49, corresponding to a 66\% relative increase. Under 100\% corruption, where \(\mathcal{G}\) is close to pure noise, the MAE reaches 30.38, which is about four times that of the uncorrupted setting. This confirms that the correctness of \(\mathcal{G}\) is important for effective meso-level sentiment. However, the degradation is not catastrophic: even under heavy corruption, the model still retains some predictive ability, likely because the AttnMLP module can down-weight the corrupted meso-level signal when it becomes too noisy.

Importantly, based on post-hoc validation, where two independent experts reviewed a 10\% random sample of entries in \(\mathcal{G}\), the agreement with the original \(\mathcal{G}\) reached 94.2\%. This suggests that the estimated error rate of our expert-constructed \(\mathcal{G}\) is approximately 6\%. We therefore conducted an additional simulation with 6\% random corruption, which yielded an MAE of 8.12, compared with 7.53 under 0\% corruption. This modest degradation, with only a \(0.59 \times 10^{-5}\) increase in MAE, indicates that our \(\mathcal{G}\) is sufficiently reliable for the intended application.

\section{Mathematical Symbol Reference}
\label{appd:symbol_ref}
As shown in Table~\ref{tab:symbol_reference}, the symbol system is organized into four groups: General Subscript, Dataset, Sentiment Modeling, and Forecasting Modeling.

\begin{table*}[!h]
\centering
\caption{A Subset of The Knowledge Graph $\mathcal{G}$.}
\label{tab:graph_exsample}
\resizebox{\textwidth}{!}{%
\begin{tabular}{lcccccc}
\toprule  
\textbf{Industry} & \makecell{\textbf{Natural}\\\textbf{Disasters}} & \makecell{\textbf{Disaster}\\\textbf{Prevention}} & 
\makecell{\textbf{Disaster}\\\textbf{Expenditure}} & 
\makecell{\textbf{Disaster}\\\textbf{Relief}} & 
\makecell{\textbf{Disaster}\\\textbf{Loss}} & 
\makecell{\textbf{Post-Disaster}\\\textbf{Reconstruction}} \\
\midrule
Agriculture, Forestry, Livestock, and Fishery & 1 & 1 & 1 & 1 & 1 & 1 \\
Basic Chemicals & 1 & 1 & 1 & 1 & 1 & 1 \\
Steel & 1 & 1 & 1 & 0 & 1 & 1 \\
Non-ferrous Metals & 1 & 1 & 1 & 0 & 1 & 1 \\
Electronics & 1 & 1 & 1 & 1 & 1 & 1 \\
Automobile & 1 & 1 & 1 & 1 & 1 & 1 \\
Household Appliances & 1 & 1 & 1 & 1 & 1 & 1 \\
Food and Beverage & 1 & 1 & 1 & 1 & 1 & 1 \\
Textiles and Apparel & 1 & 1 & 1 & 1 & 1 & 1 \\
Light Industry Manufacturing & 1 & 1 & 1 & 1 & 1 & 1 \\
Pharmaceuticals and Biotechnology & 1 & 1 & 1 & 1 & 1 & 1 \\
Utilities & 1 & 1 & 1 & 1 & 1 & 1 \\
Transportation & 1 & 1 & 1 & 1 & 1 & 1 \\
Real Estate & 1 & 1 & 1 & 0 & 1 & 1 \\
Trade and Retail & 1 & 1 & 1 & 1 & 1 & 1 \\
Tourism and Scenic Areas & 1 & 1 & 1 & 0 & 1 & 1 \\
Education (Including Sports) & 1 & 1 & 1 & 1 & 1 & 1 \\
Local Life Services & 1 & 1 & 1 & 1 & 1 & 1 \\
Professional Services & 0 & 0 & 0 & 0 & 0 & 0 \\
Hospitality and Catering & 1 & 1 & 1 & 1 & 1 & 1 \\
Banking & 0 & 0 & 0 & 0 & 0 & 0 \\
Non-bank Financial Services & 0 & 0 & 0 & 0 & 0 & 0 \\
Building Materials & 1 & 1 & 1 & 1 & 1 & 1 \\
Building Decoration & 1 & 1 & 1 & 1 & 1 & 1 \\
Electrical Equipment & 1 & 1 & 1 & 1 & 1 & 1 \\
Machinery and Equipment & 1 & 1 & 1 & 1 & 1 & 1 \\
Defense and Military Industry & 0 & 0 & 0 & 1 & 0 & 0 \\
Computer & 1 & 1 & 1 & 1 & 1 & 1 \\
Television and Broadcasting & 1 & 1 & 1 & 1 & 1 & 1 \\
Gaming & 0 & 0 & 0 & 0 & 0 & 0 \\
Advertising and Marketing & 0 & 0 & 0 & 0 & 0 & 0 \\
Film and Cinema & 1 & 1 & 0 & 0 & 1 & 0 \\
Digital Media & 0 & 0 & 0 & 0 & 0 & 0 \\
Social Media & 0 & 0 & 0 & 0 & 0 & 0 \\
Publishing & 1 & 1 & 0 & 0 & 1 & 0 \\
Telecommunications & 1 & 1 & 1 & 1 & 1 & 1 \\
Coal & 1 & 1 & 0 & 0 & 1 & 0 \\
Petroleum and Petrochemicals & 1 & 1 & 0 & 1 & 1 & 0 \\
Environmental Protection & 1 & 1 & 1 & 1 & 1 & 1 \\
Beauty and Personal Care & 0 & 0 & 0 & 0 & 0 & 0 \\
\bottomrule
\end{tabular}%
}
\footnotesize 
\begin{itemize}
\item $1$ represents that topic $n$ has an impact on industry $m$, while $0$ indicates no impact.  
\end{itemize}
\begin{tcolorbox}
\scriptsize
\begin{justify}
Natural disaster news refers to reports on the occurrence, impact, response, and recovery of natural disasters, aiming to provide timely and accurate information to help readers understand the current situation and effects of disasters. Its content includes basic information about the disaster, cause analysis, and warning information, as well as the loss of life, property, social and economic impacts, government and social rescue actions, post-disaster reconstruction, early warning systems, and emergency plans. It also covers meteorological and geological analysis, historical data and real-time data, major events, international responses, media reports, public opinions, expert analysis, and ethical controversies. Natural disaster news is not only important for general readers but also provides valuable references for policymakers, international organizations, and others. Accuracy and timeliness are crucial.
\end{justify}
\end{tcolorbox}

\end{table*}

\begin{table*}[!h]
\centering
\caption{Mathematical Symbol Reference.}
\label{tab:symbol_reference}
\small
\setlength\tabcolsep{6pt}
\captionsetup{width=\textwidth}
\begin{tabularx}{\textwidth}{p{0.13\textwidth}p{0.24\textwidth}X}
\toprule
\textbf{Category} & \textbf{Mathematical Symbol} & \textbf{Explanation} \\
\midrule

\multirow{11}{*}{\makecell[l]{General\\Subscript}} 
 & $\alpha$ & Firm-specific (micro-level) view. \\
 & $\beta$ & Industry-specific (meso-level) view. \\
 & $\text{neg}$ & The sentiment polarity ``negative''. \\
 & $\text{neu}$ & The sentiment polarity ``neutral''. \\
 & $\text{pos}$ & The sentiment polarity ``positive''. \\
 & $m$ & An arbitrary industry. \\
 & $n$ & An arbitrary topic. \\
 & $i$ & An arbitrary bond (entity). \\
 & $j$ & An arbitrary text. \\
 & $k$ & An arbitrary day. \\
 & $r$ & An arbitrary token position in text $j$. \\
\midrule
\multirow{5}{*}{\makecell[l]{Dataset}} 
 & $\mathcal{D}_{1}$ & Labeled sentiment corpus for micro-level ABSA model training. \\
 & $\mathcal{D}_{2}$ & Large-scale unlabeled corpus for market-wide sentiment inference. \\
 & $\mathcal{D}_{2,\alpha}$ & Subset of $\mathcal{D}_{2}$ used for micro-level ABSA inference. \\
 & $\mathcal{D}_{2,\beta}$ & Subset of $\mathcal{D}_{2}$ used for meso-level SLSA inference. \\
 & $\mathcal{D}_{3}$ & Bond dataset with structured features for BDRF forecasting. \\
\midrule
\multirow{37}{*}{\makecell[l]{Sentiment\\Modeling}} 
 & $s$ & Continuous sentiment score in $[-1,1]$. \\
 & $\mathbf{p}_s,\hat{\mathbf{p}}_{s}$ & Sentiment probability vector for ABSA labeling/training. \\
 & $\mathcal{G}$ & Topic--industry knowledge graph. \\
 & $g_{m,n}$ & Entry in $\mathcal{G}$; equals $1$ if topic $n$ affects industry $m$, otherwise $0$. \\
 & $\mathcal{B}$ & The definition of topics knowledge base. \\
 & $\mathcal{B}'$ & The embedding-indexed representation of $\mathcal{B}$ for retrieval. \\
 & $s_{\alpha,i,j,k}$ & Micro-level sentiment for text $j$ about bond $i$ on day $k$. \\
 & $s_{\alpha,i,k}$ & Daily micro-level sentiment of bond $i$ after averaging over texts. \\
 & $\mathcal{S}_{\alpha}$ & Matrix of firm-specific sentiment across bonds and days. \\
 & $s_{\beta,n,j,k}$ & Topic-level sentiment contribution for topic $n$ from text $j$ on day $k$. \\
 & $c_{j,n}$ & Cosine similarity between text $j$ and retrieved topic $n$ embedding. \\
 & $N^{\ast}$ & Top-$5$ retrieved topics used for meso-level mapping. \\
 & $M^{\ast}$ & Number of relevant industries linked to bond $i$. \\
 & $s_{\beta,m,k}$ & Industry-level sentiment of industry $m$ on day $k$. \\
 & $s_{\beta,i,k}$ & Bond-aligned meso-level sentiment obtained from mapped $s_{\beta,m,k}$. \\
 & $\mathcal{S}_{\beta}$ & Matrix of industry-specific sentiment across industries and days. \\
 & $\hat{s}_{\alpha,i,k}$, $\hat{s}_{\beta,i,k}$ & Input micro/meso sentiment streams fed to the aggregation module. \\
 & $\hat{s}_{i,k}$ & Aggregated sentiment of AttnMLP before duration modeling. \\
 & $h(\cdot)$ & Duration function. \\
 & $s_{i,k}$ & Final composite sentiment for bond $i$ at time step $k$. \\
 & $I, M, K$ & Number of firms, industries, and time steps (days), respectively. \\
 & $U$ & The number of texts in $\mathcal{D}_{1}$. \\
 & $J_i$ & Number of texts associated with bond $i$ on a given day. \\
 & $f_{1}, f_{2}, f_{3}$ & ABSA model, LLM agent, and embedding model, respectively. \\
 & $|l_1|,|l_2|$ & The lengths of strings $l_1$ and $l_2$.\\
 & $\tau$ & The number of matching characters between strings $l_1$ and $l_2$.\\
 & $\eta$ & Half the number of transpositions. \\
 & $l_p$ & The length of the common prefix of two strings. \\
 & $\lambda$ & The scaling coefficient controlling the contribution of  $l_p$. \\
 & $TP_{s}, FP_{s}$ & The numbers of correct/incorrect polarities in the $\mathcal{D}_{1}$ testset. \\
 & $c_{\alpha}(\cdot)$ & Micro-level sentiment classifier used after representation pooling. \\
 & $L_j$ & Number of tokens in text $j$. \\
 & $\mathbf{e}^{(j)}_{\mathrm{cls}}$ & Global text representation of text $j$ produced by the special classification token in $f_{1}$. \\
 & $\mathbf{e}^{(j)}_{r}$ & Token representation at position $r$ in text $j$ output by $f_{1}$. \\
 & $\mathcal{P}_{i,j}$ & Set of token positions corresponding to the mention of bond $i$ in text $j$. \\
 & $\mathcal{E}_{i,j}$ & Set of token representations of bond $i$ collected from text $j$. \\
 & $\mathbf{e}^{\mathrm{mean}}_{i,j}, \mathbf{e}^{\mathrm{max}}_{i,j}$ & Mean-pooled and max-pooled representations of bond $i$ in text $j$, respectively. \\

\end{tabularx}
\end{table*}

\begin{table*}[!h]
\centering
\small
\setlength\tabcolsep{6pt}
\captionsetup{width=\textwidth}
\begin{tabularx}{\textwidth}{p{0.13\textwidth}p{0.24\textwidth}X}

\multirow{8}{*}{\makecell[l]{Sentiment\\Modeling\\(cont.)}}

 & $\mathbf{z}^{\mathrm{mmp}}_{i,j}$ & Concatenated representation used for micro-level sentiment classification in mean-max pooling. \\
 & $s^{\mathrm{mmp}}_{\alpha,i,j}$ & Text-level micro sentiment score of bond $i$ in text $j$ produced by mean-max pooling. \\
 & $s^{\mathrm{mmp}}_{\alpha,i}$ & Final bond-specific micro sentiment obtained by max aggregation over relevant texts. \\
 & $\psi_{\beta}(\cdot)$ & Weighting function that maps retrieved topics to a text-level meso aggregation weight. \\
 & $s^{\mathrm{rag}}_{\beta}$ & Corpus-level meso sentiment score produced by the RAG-based mapping procedure. \\
 & $y_{j}$ & Sentiment score of text $j$ predicted by the LLM agent $f_{2}$. \\
 & $\mathbf{v}_{j}, \mathbf{v}_{n}$ & Embedding vectors of text $j$ and topic $n$ produced by $f_{3}$, respectively. \\
 & $a_{j}$ & Aggregation weight assigned to text $j$ based on the retrieved topic set. \\
\midrule
\multirow{10}{*}{\makecell[l]{Forecasting\\Modeling}}
 & $f_{4}$ & Forecasting model used in BDRF. \\
 & $f^{\ast}$ & Final integrated forecasting system in the full setting. \\
 & $\mathbf{x}_{i,k}$ & Feature vector of bond $i$ at time step $k$ in the forecasting dataset. \\
 & $\mathbf{X}_i$ & Feature set (including sentiment) for bond $i$ in BDRF. \\
 & $\mathbf{X}^{(T)}_{i,k}$ & Rolling window of length $T$ constructed from the feature sequence of bond $i$ starting at time step $k$. \\
 & $\mathbf{y}_i$, $\hat{\mathbf{y}}_i$ & Ground-truth and forecasted credit spread sequences for bond $i$. \\
 & $\hat{y}_{i,k+q}$ & Forecasted credit spread of bond $i$ at forecast target time step $k+q$. \\
 & $N_{train}, N_{test}$ & The total number of bonds in $\mathcal{D}_{3}$ trainset/testset.\\
 & $T$ & Rolling window length in time-series forecasting. \\
 & $q$ & Forecast horizon (the $q$-th day after the observation window). \\
\bottomrule
\end{tabularx}
\end{table*}

\end{document}